\documentclass[aps,pra,reprint,superscriptaddress,showpacs]{revtex4-1}

\usepackage{graphicx}
\usepackage{amssymb,amsmath}
\usepackage{bm}
\usepackage{dcolumn}

\begin{document}

\title{Quantum-defect theory for $-1/r^4$ type of interactions}

\author{Bo Gao}
\email[]{bo.gao@utoledo.edu}
\homepage[]{http://bgaowww.physics.utoledo.edu}
\affiliation{Department of Physics and Astronomy,
	Mailstop 111,
	University of Toledo,
	Toledo, Ohio 43606,
	USA}

\date{June 9, 2013}

\begin{abstract}

We present a quantum-defect theory (QDT) for the $-1/r^4$
type of long-range potential, as a foundation for a systematic understanding
of charge-neutral quantum systems such as ion-atom, ion-molecule, 
electron-atom, and positron-atom interactions. 
The theory incorporates both conceptual and
mathematical advances since earlier formulations of
the theory. It also includes more detailed discussions of the concept of
resonance spectrum and its representations, universal properties
in charge-neutral quantum systems,
and the QDT description of scattering resonances that is applicable 
to any $-1/r^n$ potential with $n>2$.

\end{abstract}


\maketitle

\section{Introduction}

The quantum-defect theory (QDT) for $-1/r^4$ type of interactions, 
if broadly defined as a quantum theory that explicitly takes advantage of the
universality due to the long-range potential, has existed in
various forms for decades \cite{oma61,wat80,fab86}. Notably, the
theory of O'Malley \textit{et al.} \cite{oma61} gives an analytic description
of ultracold electron-atom and ion-atom collision 
that has stood for many years. The theory of Fabrikant \cite{fab86} gives
a theory of scattering that takes further advantage of the
modified Mathieu functions \cite{hol73,khr93,olv10}.
The theory of Watanabe and Greene \cite{wat80} gives a more complete QDT formulation for
$-1/r^4$ potential in that it treats both positive and negative energies
in a consistent QDT manner which is important for, e.g., its application
in a multichannel formulation to describe Fano-Feshbach resonances.
Together, these theories have provided a solid theoretical backbone for
our understanding of charge-neutral quantum systems in low-energy regimes
or around a dissociation (detachment) threshold, 
and have served us well for many years, including in 
more recent applications
such as Rydberg molecules \cite{gre00,ben09,ben10,Low2012}.

Renewed interest in QDT for $-C_4/r^4$ polarization potential has arisen with the
emergence of cold ion-atom \cite{cot00,gri09,idz09,gao10a,zip10,zip10b,Schmid2010,rel11,hal11,idz11}
and ion-molecule interactions and reactions \cite{Willitsch2008,sta08,Roth2008,hud09,gao11a,wil12}.
For such heavier systems, the QDT takes on a different magnitude of
importance for three primary reasons. First, a same 
long-range polarization potential, 
which in the case of electron-atom interaction can only bind a
few states or lead to a few resonances \cite{buc94}, can bind many more states
and lead to many more resonances.
Their existence implies a rapid energy variation induced by the
long-range interaction, making the QDT description far more
important and necessary. Mathematically, this greater importance
of long-range potential for heavier systems is reflected in the
length scale, $\beta_4 \equiv (2\mu C_4/\hbar^2)^{1/2}$, 
and the corresponding energy scale, $s_E = (\hbar^2/2\mu)(1/\beta_4)^2$,
associated with a $-C_4/r^4$ interaction.
The length scale $\beta_4$ scales with the
reduced mass as $\mu^{1/2}$, and is hundreds times greater for
ion-atom \cite{gao10a} and ion-molecule \cite{gao11a} systems 
than for electron-atom systems. 
The energy scale $s_E$ scales with the reduced mass as $\mu^{-2}$, and
is typically $10^{9}$ times smaller \cite{gao10a,gao11a}. Between a fixed energy $\epsilon$
and the dissociation threshold $\epsilon=0$, the number of bound states or resonances 
due to the long-range potential is determined by the scaled energy $\epsilon/s_E$ 
(see Ref.~\cite{gao10a} and Sec.~\ref{sec:specreps}), which is vastly greater for
heavier systems.
Second, at any fixed positive energy $\epsilon>0$ above the threshold, 
vastly many more partial 
waves, of the order of $\sqrt{2}(\epsilon/s_E)^{1/4}$, contribute to
ion-atom and ion-molecule interactions than to electron-atom interactions.
At room temperature, e.g., hundreds of partial waves contribute to typical
ion-atom scattering, while only one or a few to typical electron-atom scattering.
An efficient and unified description of a large number of partial 
waves \cite{gao01,gao08a} is thus critically important to any systematic
quantum theory of heavier charge-neutral systems that intends to cover a wide range 
of energies such as from absolute zero temperature to the room temperature.
We emphasize that large number of partial waves and seemingly small de Broglie
wave length do not guarantee classical behavior, there are subtle quantum
effects such as shape resonances that can persist even when such conditions
seem well satisfied. These ``high temperature'' resonances can be
expected to play an important role in chemistry such as molecule 
formation in a dilute environment \cite{cha10,bar06}, and in thermodynamics.
Third, ion-atom and ion-molecule interactions are extremely sensitive
to the short-range potential \cite{hec02,LG12}, due again to their large reduced mass \cite{gao96}.
With the exception of H$^+$+H and 
its isotopic variations \cite{iga99,esr00,Kristic2004,bod08},
this sensitive dependence makes most theoretical predictions based on
\textit{ab initio} potentials unreliable. 
QDT and related multichannel quantum-defect theory (MQDT), 
especially though their partial-wave
insensitive formulations \cite{gao01,gao05a}, offer a prospect to overcome
this difficulty by reducing their description to very few parameters
that can be determined with a few experimental data points 
\cite{car88,car93,car95a,car95b,hec02},
without relying on the precise knowledge of the short-range potential \cite{LG12}.

We present here a version of the QDT for the $-1/r^4$ potential that
incorporates both recent conceptual advances in QDT \cite{gao08a,gao10a}
and mathematical advances in the understanding of the modified Mathieu 
functions \cite{gao10a,idz11}. A brief account of the theory, and its
initial applications to ion-atom interactions and charge-neutral
reactions, have been presented in Refs.~\cite{gao10a,gao11a,LG12}.
This work presents the details of the underlying QDT formulation,
in preparation for its further applications.
It includes derivations of the quantum reflection and transmission amplitudes
for the $-1/r^4$ potential (important for understanding charge-neutral reactive
processes \cite{gao11a}), a more detailed discussion of the concept of
resonance spectrum \cite{gao10a} and its representations, universal properties
in charge-neutral quantum systems especially ion-atom interactions,
and the QDT description of scattering resonances that is applicable 
to any $-1/r^n$ potential with $n>2$.
This presentation also serves as a concrete example
showing how the general QDT structure of Ref.~\cite{gao08a} 
is actually realized for a particular $n$. 

We begin by recalling the keys steps in constructing a QDT
for a $-C_n/r^n$ ($n>2$) potential \cite{gao08a}.
We want to first find solutions of the Schr\"{o}dinger equation
\begin{equation}
\left[-\frac{\hbar^2}{2\mu}\frac{d^2}{dr^2}+\frac{\hbar^2l(l+1)}{2\mu r^2}
	- \frac{C_n}{r^n}-\epsilon\right]
	v_{\epsilon l}(r) = 0 \;,
\label{eq:invrnsch}
\end{equation}
where $\mu$ is the reduced mass, and $\epsilon$ is the energy. 
After scaling the length by the length scale
\begin{equation}
\beta_n \equiv (2\mu C_n/\hbar^2)^{1/(n-2)} \;,
\label{eq:betan}
\end{equation}
and the energy by a corresponding energy scale $s_E = (\hbar^2/2\mu)(1/\beta_n)^2$.
Eq.~(\ref{eq:invrnsch}) takes the dimensionless
form of 
\begin{equation}
\left[\frac{d^2}{dr_s^2} - \frac{l(l+1)}{r_s^2}
	+ \frac{1}{r_s^n} + \epsilon_s\right]
	v_{\epsilon_s l}(r_s) = 0 \;,
\label{eq:invrnschs}
\end{equation}
where $r_s = r/\beta_n$ is a scaled radius, and $\epsilon_s = \epsilon/s_E$
is a scaled energy. 
We would like to find a pair of linearly independent solutions
of Eq.~(\ref{eq:invrnschs}),
the so-called QDT base pair \cite{gao08a}, defined by
the small-$r_s$ asymptotic behavior of
\begin{eqnarray}
f^c_{\epsilon_s l}(r_s) &\stackrel{r_s\rightarrow 0}{\sim}& 
	(2/\pi)^{1/2}r_s^{n/4}
	\cos\left(y-\pi/4 \right) \;, 
\label{eq:fcdef}\\
g^c_{\epsilon_s l}(r_s) &\stackrel{r_s\rightarrow 0}{\sim}& 
	-(2/\pi)^{1/2}r_s^{n/4}
	\sin\left(y -\pi/4 \right) \;,
\label{eq:gcdef}
\end{eqnarray}
for all energies. Here $y=[2/(n-2)]r_s^{-(n-2)/2}$.

The large-$r_s$ asymptotic behaviors of such an pair, in the limit
of $r_s\rightarrow\infty$, defines the $Z^c$ matrix for 
positive energies, and the $W^c$ matrix for negative energies,
the combination of which gives one formulation of QDT 
for $-1/r^n$ potential \cite{gao08a}.
From the $Z^c$ and $W^c$ matrices, one can derive the
quantum reflection and transmission amplitudes associated
with the long-range potential, from which a different
QDT formulation can be constructed \cite{gao08a}.
This latter formulation, namely QDT in terms of
reflection and transmission amplitudes, especially
its multichannel generalization \cite{gao10b},
is playing an important role in applications of
QDT in reactions and inelastic processes \cite{gao10b,gao11a}.

For $n=4$, the solutions of Eq.~(\ref{eq:invrnschs}) are given 
in terms of the modified Mathieu functions \cite{hol73,khr93,olv10}. 
While they are well-known mathematical special functions, their
understanding and application in physics 
have been somewhat limited by their relative complexity.
Our QDT for $-1/r^4$ potential includes an alternative
method of solving and understanding Mathieu class of 
functions that may help to stimulate their further applications.
The method further emphasizes the structural similarities of
the $1/r^4$ solutions to solutions for $1/r^6$ \cite{gao98a} and 
$1/r^3$ \cite{gao99a} potentials, which should be helpful
in understanding all such solutions.

The paper is organized as follows. In Sec.~\ref{sec:qdtfuncs},
we present the QDT functions for $-1/r^4$ potential,
including the reference wave functions, 
the $Z^c$ and $W^c$ matrices,
the quantum reflection and transmission amplitudes,
and the corresponding QDT functions for negative energies,
such as the quantum order parameter introduced in Ref.~\cite{gao08a}.
The key results of the corresponding single-channel QDT 
for $-1/r^4$ interaction are presented in Sec.~\ref{sec:qdt}.
It includes a unified understanding of both the bound spectrum 
and the resonance spectrum, their different representations,
and a QDT description of scattering resonance that is
applicable to any $-1/r^n$ potential with $n>2$.
Section~\ref{sec:ub4} discusses the single-channel universal behaviors 
for charge-neutral quantum systems, especially ion-atom systems,
as implied in the QDT formulation.
In Sec.~\ref{sec:discuss}, we briefly discuss the
differences in applying the theory to ion-atom and
to electron-atom interactions.
Section~\ref{sec:conclusion} concludes the article.

\section{QDT functions for $-1/r^4$ potential}
\label{sec:qdtfuncs}

\subsection{The math reference pair}

Specializing to the $-C_4/r^4$ potential
with $C_4>0$, the length scale $\beta_n$ becomes 
$\beta_4 \equiv (2\mu C_4/\hbar^2)^{1/2}$,
and Eq.~(\ref{eq:invrnschs}) becomes
\begin{equation}
\left[\frac{d^2}{dr_s^2} - \frac{l(l+1)}{r_s^2}
	+ \frac{1}{r_s^4} + \epsilon_s\right]
	v_{\epsilon_s l}(r_s) = 0 \;.
\label{eq:invr4schs}
\end{equation}
One pair of its solutions, which we call the math pair, is 
given in terms of the modified Mathieu functions \cite{hol73}
\begin{eqnarray}
\xi_{\epsilon_s l}(r_s) &=& r_s^{1/2} {\mathcal M}_{+\nu}(x) \;,
\label{eq:xi}\\
\eta_{\epsilon_s l}(r_s) &=& r_s^{1/2}{\mathcal M}_{-\nu}(x)\;.
\label{eq:eta}
\end{eqnarray}
Here $x=\epsilon_s^{1/4}r_s$, and ${\mathcal M}_{+\nu}(x)$ and 
${\mathcal M}_{-\nu}(x)$ are the modified Mathieu functions with 
Laurent expansions \cite{hol73}
\begin{eqnarray}
{\mathcal M}_{+\nu}(x) &=& \sum_{m=-\infty}^{\infty}
	b_m x^{\nu+2m} \;,
\label{eq:Mp}\\
{\mathcal M}_{-\nu}(x) &=& \sum_{m=-\infty}^{\infty}
	b_{-m} x^{-\nu+2m} \;.
\label{eq:Mm}
\end{eqnarray}
In Eqs.~(\ref{eq:Mp}) and (\ref{eq:Mm}), 
the normalization is chosen such that $b_0=1$.
The coefficients $b_j$ satisfy a set of well-known three-term
recurrence relations for Mathieu class of functions \cite{hol73}
\begin{equation}
h_mb_{m+1}+b_m+h_mb_{m-1} = 0 \;,
\label{eq:mathieurec}
\end{equation}
with
\begin{equation}
h_m = \epsilon_s^{1/2}/[(\nu+2m)^2-\nu_0^2] \;.
\label{eq:hm}
\end{equation}
Here $\nu_0=l+1/2$, and $\nu$ is the characteristic exponent for the
$-1/r^4$ potential, discussed further in the Appendix~\ref{sec:charexp4}.
We have solved this set of recurrence relations using the method
developed earlier for $1/r^6$ \cite{gao98a} and $1/r^3$
\cite{gao99a} potentials to give
\begin{eqnarray}
b_j &=& (-\Delta)^j \frac{\Gamma[1+(\nu-\nu_0)/2]
	\Gamma[1+(\nu+\nu_0)/2]}{\Gamma[j+1+(\nu-\nu_0)/2]
	\Gamma[j+1+(\nu+\nu_0)/2]}c_j(\nu) \;,
\label{eq:bpj}\\
b_{-j} &=& (-\Delta)^j \frac{\Gamma[(\nu-\nu_0)/2-j]
	\Gamma[(\nu+\nu_0)/2-j]}{\Gamma[(\nu-\nu_0)/2]
	\Gamma[(\nu+\nu_0)/2]}c_j(-\nu) \;.
\label{eq:bmj}
\end{eqnarray}
In Eqs.~(\ref{eq:bpj}) and (\ref{eq:bmj}), $j$ is a positive
integer, $\Delta = \epsilon_s^{1/2}/4$, and $\Gamma(x)$ is
the standard gamma function \cite{olv10}.
The $c_j(\nu)$ coefficients are given by
\begin{equation}
c_j(\nu) = \prod_{m=0}^{j-1}Q(\nu+2m) \;,
\label{eq:cj}
\end{equation}
in which $Q(\nu)$ is given by a continued 
fraction
\begin{equation}
Q(\nu) = \frac{1}{1-\frac{\epsilon_s}{
	[(\nu+2)^2-\nu_0^2][(\nu+4)^2-\nu_0^2]}
	Q(\nu+2)} \;.
\label{eq:Qcf4}
\end{equation}

With analytic expressions for $b_j$ as given by 
Eqs.~(\ref{eq:bpj}) and (\ref{eq:bmj}), 
the asymptotic behaviors of the math pair, for both small $r_s$
and large $r_s$, can be derived directly from their Laurent
expansions, using a method that is similar to what led to the large-$r$
behaviors of the $-1/r^6$ solutions \cite{gao98a}.
For small $r_s$, we obtain
\begin{eqnarray}
\xi_{\epsilon_s l} &\stackrel{r_s\rightarrow 0}{\sim}& 
	F_{\epsilon_s l}(-\nu) r_s^{1/2} \lim_{r_s\rightarrow 0} 
	J_{-\nu} \left(y\right)
	\nonumber \\
	& \sim &
	F_{\epsilon_s l}(-\nu)(-1)^{l+1}(2/\pi)^{1/2} r_s \nonumber\\
	& & \times\left[ \cos(\pi\nu/2)
	\cos\left( y - \pi/4 \right)
	\right.\nonumber\\
	& & -\left.\sin(\pi\nu/2)
	\sin\left( y - \pi/4 \right) 
	\right] \;,
\end{eqnarray}
\begin{eqnarray}
\eta_{\epsilon_s l} &\stackrel{r_s\rightarrow 0}{\sim}& 
	F_{\epsilon_s l}(+\nu)r_s^{1/2}\lim_{r_s\rightarrow 0} 
	J_{\nu} \left(y\right)
	\nonumber \\
	& \sim &
	F_{\epsilon_s l}(+\nu)(2/\pi)^{1/2} r_s \nonumber\\
	& & \times\left[ \cos(\pi\nu/2)
	\cos\left( y - \pi/4 \right)
	\right.\nonumber\\
	& & +\left.\sin(\pi\nu/2)
	\sin\left( y - \pi/4 \right) 
	\right] \;.
\end{eqnarray}
where $y=[2/(n-2)]r_s^{-(n-2)/2}=1/r_s$ for $n=4$,
$J_{\pm\nu}(x)$ are the Bessel functions \cite{olv10}, and
\begin{equation}
F_{\epsilon_s l}(\nu) = 2^{\nu}\epsilon_s^{-\nu/4}\Gamma[1+(\nu+\nu_0)/2]
	\Gamma[1+(\nu-\nu_0)/2]C_{\epsilon_s l}(\nu) \;,
\label{eq:Fnu}
\end{equation}
in which
\begin{equation}
C_{\epsilon_s l}(\nu) = \lim_{j\rightarrow\infty}c_j(\nu) = \prod_{j=0}^{\infty} Q(\nu+2j) \;.
\label{eq:Cnu4}
\end{equation}
For large $r_s$, the asymptotic behaviors of the math pair are given 
for $\epsilon_s>0$ by
\begin{eqnarray}
\xi_{\epsilon_s l} &\stackrel{r_s\rightarrow \infty}{\sim}& 
	F_{\epsilon_s l}(+\nu) r_s^{1/2} \lim_{r_s\rightarrow \infty} 
	J_{\nu} \left(k_sr_s\right) \nonumber \\
	& \sim &
	F_{\epsilon_s l}(+\nu)(2/\pi k_s)^{1/2}  \nonumber\\
	& & \times\left[ \cos[\pi(\nu-\nu_0)/2]
	\sin(k_sr_s - l\pi/2)
	\right.\nonumber\\
	& & -\left.\sin[\pi(\nu-\nu_0)/2]
	\cos(k_sr_s - l\pi/2) 
	\right] \;,
\label{eq:xilargerp}
\end{eqnarray}
\begin{eqnarray}
\eta_{\epsilon_s l} &\stackrel{r_s\rightarrow \infty}{\sim}& 
	F_{\epsilon_s l}(-\nu)r_s^{1/2}\lim_{r_s\rightarrow \infty} 
	J_{-\nu} \left(k_sr_s\right) \nonumber \\
	& \sim &
	(-1)^lF_{\epsilon_s l}(-\nu)(2/\pi k_s)^{1/2}  \nonumber\\
	& & \times\left[-\sin[\pi(\nu-\nu_0)/2]
	\sin(k_sr_s - l\pi/2)
	\right.\nonumber\\
	& & +\left.\cos[\pi(\nu-\nu_0)/2]
	\cos(k_sr_s - l\pi/2) 
	\right] \;,
\end{eqnarray}
where $k_s=\epsilon_s^{1/2}$,
and for $\epsilon_s<0$ by
\begin{eqnarray}
\xi_{\epsilon_s l} &\stackrel{r_s\rightarrow \infty}{\sim}& 
	F_{\epsilon_s l}(+\nu) r_s^{1/2} \lim_{r_s\rightarrow \infty} 
	I_{\nu} \left(\kappa_sr_s\right)
	\nonumber \\
	& \sim &
	F_{\epsilon_s l}(+\nu)\frac{1}{\sqrt{2\pi\kappa_s}}
	\left[-\sin(\pi\nu) e^{-\kappa_sr_s}+e^{+\kappa_sr_s}\right] \;,
\end{eqnarray}
\begin{eqnarray}
\eta_{\epsilon_s l} &\stackrel{r_s\rightarrow \infty}{\sim}& 
	F_{\epsilon_s l}(-\nu)r_s^{1/2}\lim_{r_s\rightarrow \infty} 
	I_{-\nu} \left(\kappa_sr_s\right)
	\nonumber \\
	& \sim &
	F_{\epsilon_s l}(-\nu)\frac{1}{\sqrt{2\pi\kappa_s}}
	\left[\sin(\pi\nu) e^{-\kappa_sr_s}+e^{\kappa_sr_s}\right] \;,
\label{eq:etalargern}
\end{eqnarray}
where $\kappa_s=(-\epsilon_s)^{1/2}$, and $I_{\pm\nu}(x)$ are the
modified Bessel functions \cite{olv10}.
An equivalent pair of solutions has been found independently by
Idziaszek \textit{et al.} \cite{idz11}, using a similar
method \cite{gao98a,gao99a}.

\subsection{The QDT base pair and the $Z^c$ and $W^c$ matrices}

The QDT base pair, $f^c$ and $g^c$, has been defined in 
a way that they have energy and partial wave independent asymptotic behaviors in 
the region of $r\ll\beta_4$ ($r_s\ll 1$), 
given by [c.f. Eqs.~(\ref{eq:fcdef}) and (\ref{eq:gcdef})]
\begin{eqnarray}
f^c_{\epsilon_s l}(r_s) &\stackrel{r_s\ll 1}{\sim}& 
	(2/\pi)^{1/2}r_s
	\cos\left(y-\pi/4 \right) \;, 
\label{eq:fcdef4}\\
g^c_{\epsilon_s l}(r_s) &\stackrel{r_s\ll 1}{\sim}& 
	-(2/\pi)^{1/2}r_s
	\sin\left(y -\pi/4 \right) \;,
\label{eq:gcdef4}
\end{eqnarray}
for all energies \cite{gao01,gao08a}. Here $y=1/r_s$ as defined earlier.
They are normalized such that 
\begin{equation}
W(f^c,g^c) \equiv f^c\frac{d g^c}{dr_s}-\frac{d f^c}{dr_s}g^c
	= 2/\pi \;.
\end{equation}

From the definitions of $f^c$ and $g^c$, and the small-$r_s$ asymptotic
behaviors of the math pair, it is straightforward to show that
the QDT base pair is given in terms of the math pair by
\begin{eqnarray}
f^c_{\epsilon_s l}(r) &=& \frac{1}{2\cos(\pi\nu/2)}
	\left[\frac{1}{F_{\epsilon_s l}(-\nu)}\xi_{\epsilon_s l}
	+ \frac{1}{F_{\epsilon_s l}(+\nu)}\eta_{\epsilon_s l}\right] ,
\label{eq:fc}\\
g^c_{\epsilon_s l}(r) &=& \frac{1}{2\sin(\pi\nu/2)}
	\left[\frac{1}{F_{\epsilon_s l}(-\nu)}\xi_{\epsilon_s l}
	- \frac{1}{F_{\epsilon_s l}(+\nu)}\eta_{\epsilon_s l}\right] \;.
\label{eq:gc}
\end{eqnarray}
From the solutions for $\xi_{\epsilon_s l}$, $\eta_{\epsilon_s l}$,
and the definition of $F_{\epsilon_s l}(+\nu)$, one can verify that
$\xi_{\epsilon_s l}/F_{\epsilon_s l}(-\nu)$, 
$\eta_{\epsilon_s l}/F_{\epsilon_s l}(+\nu)$, and hence
$f^c_{\epsilon_s l}$ and $g^c_{\epsilon_s l}$, are entire functions
of $\epsilon_s$.
Physically, this is what ensures that the short-range $K^c$ matrix,
defined in reference to the QDT base pair,
being meromorphic in energy \cite{gao08a}.
Mathematically, it allows analytic continuation of the base pair to
negative energies (and complex energies if necessary) without
explicitly solutions of the math pair for such
energies.

The large-$r_s$ asymptotic behaviors of the QDT base pair,
which give the $Z^c$ and the $W^c$ matrices, follow
from Eqs.~(\ref{eq:fc}) and (\ref{eq:gc}), and the
large-$r_s$ behaviors of the math pair, as given by
Eqs.~(\ref{eq:xilargerp})-(\ref{eq:etalargern}).
For $\epsilon_s>0$, we obtain
\begin{eqnarray}
f^c_{\epsilon_s l}(r_s) &\stackrel{r_s\rightarrow \infty}{\sim}&
	\left(\frac{2}{\pi k_s}\right)^{1/2} \left[Z^c_{fs}
	\sin\left(k_sr_s-\frac{l\pi}{2}\right) \right.\nonumber\\
	& &\left. -Z^c_{fc}\cos\left(k_sr_s-\frac{l\pi}{2}\right)\right] ,
\label{eq:fclarger}\\	 
g^c_{\epsilon_s l}(r_s) &\stackrel{r_s\rightarrow \infty}{\sim}&
	\sqrt{\frac{2}{\pi k_s}} \left[Z^c_{gs}
	\sin\left(k_sr_s-\frac{l\pi}{2}\right) \right.\nonumber\\
	& &\left. -Z^c_{gc}\cos\left(k_sr_s-\frac{l\pi}{2}\right)\right] ,
\label{eq:gclarger}
\end{eqnarray}
with
\begin{eqnarray}
Z^c_{fs}(\epsilon_s,l) &=& 
	\frac{\cos[\pi(\nu-\nu_0)/2]}
	{2M_{\epsilon_s l}\cos(\pi\nu/2)} \nonumber\\
	& &\times\left\{1-(-1)^l M_{\epsilon_s l}^2\tan[\pi(\nu-\nu_0)/2]\right\} \;, \\
Z^c_{fc}(\epsilon_s,l) &=& 
	\frac{\cos[\pi(\nu-\nu_0)/2]}
	{2M_{\epsilon_s l}\cos(\pi\nu/2)} \nonumber\\
	& &\times\left\{\tan[\pi(\nu-\nu_0)/2]-(-1)^l M_{\epsilon_s l}^2\right\} \;, \\
Z^c_{gs}(\epsilon_s,l) &=& 
	\frac{\cos[\pi(\nu-\nu_0)/2]}
	{2M_{\epsilon_s l}\sin(\pi\nu/2)} \nonumber\\
	& &\times\left\{1+(-1)^l M_{\epsilon_s l}^2\tan[\pi(\nu-\nu_0)/2]\right\} \;, \\
Z^c_{gc}(\epsilon_s,l) &=& 
	\frac{\cos[\pi(\nu-\nu_0)/2]}
	{2M_{\epsilon_s l}\sin(\pi\nu/2)} \nonumber\\
	& &\times\left\{\tan[\pi(\nu-\nu_0)/2]+(-1)^l M_{\epsilon_s l}^2\right\} \;.
\end{eqnarray}
For $\epsilon_s<0$, we obtain
\begin{eqnarray}
f^c_{\epsilon_s l}(r_s) &\stackrel{r_s\rightarrow \infty}{\sim}&
	\frac{1}{\sqrt{\pi\kappa_s}}\left[W^c_{f+}e^{-\kappa_s r_s}
	+W^c_{f-}e^{+\kappa_s r_s}\right] \;,
\label{eq:fclrne}\\	 
g^c_{\epsilon_s l}(r_s) &\stackrel{r_s\rightarrow \infty}{\sim}&
	\frac{1}{\sqrt{\pi\kappa_s}}\left[W^c_{g+}e^{-\kappa_s r_s}
	+W^c_{g-}e^{+\kappa_s r_s}\right] \;,
\label{eq:gclrne}
\end{eqnarray}
with 
\begin{eqnarray}
W^c_{f+}(\epsilon_s,l) &=& -\frac{\sin(\pi\nu/2)(1-M_{\epsilon_s l}^2)}
	{2^{1/2}M_{\epsilon_s l}} ,\\
W^c_{f-}(\epsilon_s,l) &=& \frac{1+M_{\epsilon_s l}^2}
	{2^{3/2}M_{\epsilon_s l}\cos(\pi\nu/2)} ,\\
W^c_{g+}(\epsilon_s,l) &=& -\frac{\cos(\pi\nu/2)(1+M_{\epsilon_s l}^2)}
	{2^{1/2}M_{\epsilon_s l}} ,\\
W^c_{g-}(\epsilon_s,l) &=& \frac{1-M_{\epsilon_s l}^2}
	{2^{3/2}M_{\epsilon_s l}\sin(\pi\nu/2)} \;.
\end{eqnarray}
In the expressions for the $Z^c$ and $W^c$ matrices,
we have used the definition
\begin{equation}
G_{\epsilon_s l}(\nu) \equiv 2^{\nu}|\epsilon_s|^{-\nu/4}\Gamma[1+(\nu+\nu_0)/2]
	\Gamma[1+(\nu-\nu_0)/2]C_{\epsilon_s l}(\nu) \;,
\label{eq:Gnu}
\end{equation}
to define
\begin{eqnarray}
M_{\epsilon_s l}(\nu) &\equiv& G_{\epsilon_s l}(-\nu)/G_{\epsilon_s l}(+\nu) \;,\\
	&=& 2^{-2\nu}|\epsilon_s|^{\nu/2}
	\left(\frac{\Gamma[1-(\nu+\nu_0)/2]}{\Gamma[1+(\nu+\nu_0)/2]}\right) \nonumber\\
	& &\times \left(\frac{\Gamma[1-(\nu-\nu_0)/2]}{\Gamma[1+(\nu-\nu_0)/2]}\right)
	\left(\frac{C_{\epsilon_s l}(-\nu)}{C_{\epsilon_s l}(+\nu)}\right) \;,
\label{eq:Mnu4}
\end{eqnarray}
for both the positive and the negative energies.
We note the subtle difference between the $G_{\epsilon_s l}(\nu)$,
defined by Eq.~(\ref{eq:Gnu}), and the $F_{\epsilon_s l}(\nu)$,
defined by Eq.~(\ref{eq:Fnu}), which is the result of careful
analytic continuation through entire functions \footnote{The negative energy solutions have also
been independently verified through explicit solutions of Eq.~(\ref{eq:invr4schs})
for negative energies.}.

\begin{figure}
\includegraphics[width=\columnwidth]{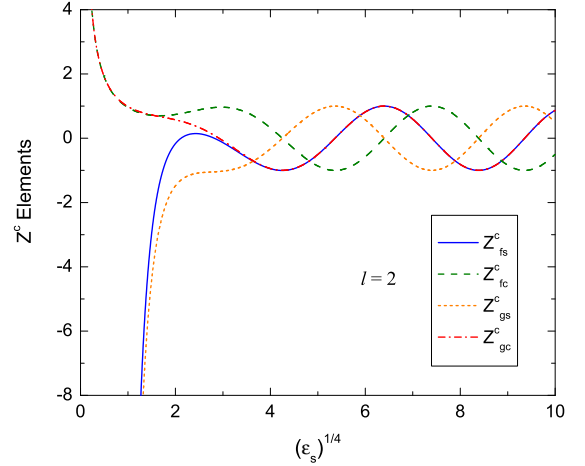}
\caption{(Color online) The (dimensionless) $Z^c$ matrix elements for the $-1/r^4$ 
potential and $l=2$. They, as are all QDT functions, are functions of 
a (dimensionless) scaled energy. The $Z^c$ matrix is defined only for
positive energies, and is presented here on the natural energy scale of
$\epsilon_s^{1/4}$ for the $-1/r^4$ potential.
\label{fig:Zc2}}
\end{figure}
\begin{figure}
\includegraphics[width=\columnwidth]{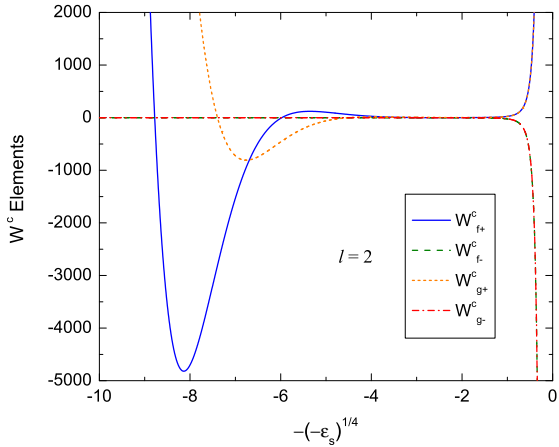}
\caption{(Color online) The (dimensionless) $W^c$ matrix elements for the $-1/r^4$ 
potential and $l=2$. The $W^c$ matrix is 
defined only for negative energies, and is presented here on the natural energy scale of
$-(-\epsilon_s)^{1/4}$ for the $-1/r^4$ potential.
\label{fig:Wc2}}
\end{figure}

The $M_{\epsilon_s l}$ function, which is function of the scaled energy $\epsilon_s$,
is one of the most important mathematical entities in QDT for the $-1/r^4$ potential.
All QDT functions of physical interest for
the $-1/r^4$ potential can be represented in terms of $M_{\epsilon_s l}$ and the
characteristic exponent $\nu$, which is itself a function of $\epsilon_s$ 
(see Appendix~\ref{sec:charexp4}). Furthermore, all singular behaviors
at $\epsilon_s=0$ are isolated to the $|\epsilon_s|^{\nu/2}$ factor within
$M_{\epsilon_s l}$.

The $Z^c$ and $W^c$ matrices, which are constrained by
$\det(Z^c)=1$ and $\det(W^c)=1$, give one formulation of
the QDT for $-1/r^4$ potential \cite{gao08a,gao09a}, 
to be discussed further in Sec.~\ref{sec:qdt}.
We note that the determinant constraints on $Z^c$
and $W^c$ are automatically ensured by their
representations in terms of $M_{\epsilon_s l}$ and $\nu$.

The elements of $Z^c$ and $W^c$ matrices are illustrated for $l=2$ in 
Figs.~\ref{fig:Zc2} and \ref{fig:Wc2}, respectively, on the natural energy scales of
$\epsilon_s^{1/4}$ for positive energies and $-(-\epsilon_s)^{1/4}$ for
negative energies. More generally for a $-1/r^n$ potential, there
exist natural energy scales of $\epsilon_s^{(n-2)/2n}$ for positive energies 
and $-(-\epsilon_s)^{(n-2)/2n}$ for negative energies.
They are energy scales associated with the semiclassical behaviors away 
from the threshold \cite{ler70,gao99b,fla99,Friedrich2004}.
In later figures covering both positive and negative energies,
the natural energy scale for the $-1/r^4$ potential will be represented as
$\mathrm{sgn}(\epsilon_s)|\epsilon_s|^{1/4}$, with $\mathrm{sgn}(\epsilon_s)$
defined by
\[
\mathrm{sgn}(\epsilon_s) = \left\{
\begin{array}{ll}
-1 & \epsilon_s<0 \;,\\
0 &  \epsilon_s=0 \;,\\
+1 & \epsilon_s>0 \;.
\end{array}
\right.
\]

\subsection{Quantum reflection and transmission amplitudes}
\label{sec:qrefl}

Instead of the $Z^c$ matrices, QDT for positive energies can also be
constructed in terms of the quantum reflection and transmission amplitudes
associated with the long-range potential \cite{gao08a}.
Such a formulation, especially its multichannel generalization \cite{gao10b},
has clearer physical interpretation and
has proven to be especially effective in treating and 
understanding reactive and inelastic processes \cite{gao10b,gao11a}.

There are four such amplitudes for each partial wave $l$.
The two for reflection by the long-range potential
can be written as \cite{gao08a}
\begin{eqnarray}
r^{(oi)}_l &=& (-1)^l\sqrt{{\mathcal R}^c_l}\exp\left[i(\delta^{c}_l+\phi^{c}_l)\right] \;,
\label{eq:roi} \\
r^{(io)}_l &=& \sqrt{{\mathcal R}^c_l}\exp\left[i(\delta^{c}_l-\phi^{c}_l)\right] \;,
\label{eq:rio}
\end{eqnarray}
where $r^{(oi)}_l$ and $r^{(io)}_l$ represent the reflection amplitudes
by the long-range potential for particles going outside-in (approaching each other)
and inside-out (moving away from each other), respectively.
The two amplitudes for transmission can be written as \cite{gao08a}
\begin{equation}
t^{(io)}_l = t^{(oi)}_l = 
\sqrt{{\mathcal T}^c_l}\exp\left(-il\pi/2-i\pi/2+i\delta^{c}_l\right) \;.
\label{eq:tl}
\end{equation}
where $t^{(oi)}_l$ and $t^{(io)}_l$ represent the transmission amplitudes
through the long-range potential for particles going outside-in 
and inside-out, respectively.
Equations~(\ref{eq:roi})-(\ref{eq:tl}) imply that the two transmission
amplitudes are always equal, while the two reflection amplitudes generally
differ by a phase. All amplitudes can be determined from
three independent functions: (a) the quantum reflection probability
${\mathcal R}^c_l$ or the quantum transmission probability 
${\mathcal T}^c_l$, which are related by 
${\mathcal T}^c_l=1-{\mathcal R}^c_l$, (b) the long-range (transmission)
phase shift $\delta^{c}_l$, and (c) the reflection phase shift
$\phi^{c}_l$, all of which can be determined from 
the $Z^c$ matrix \cite{gao08a}.

From the $Z^c$ matrix of the previous section,
we obtain for $-1/r^4$ potential the quantum reflection probability
\begin{eqnarray}
{\mathcal R}^c_l(\epsilon_s) &=& \frac{(Z^c_{fs}-Z^c_{gc})^2+(Z^c_{fc}+Z^c_{gs})^2}
	{(Z^c_{fs}+Z^c_{gc})^2+(Z^c_{fc}-Z^c_{gs})^2}\;, \\
	&=& \frac{(1-M_{\epsilon_s l}^2)^2}{1-2M_{\epsilon_s l}^2\cos(2\pi\nu)+M_{\epsilon_s l}^4} \;.
\label{eq:rp4}	
\end{eqnarray}
The related transmission probability is given by
\begin{eqnarray}
{\mathcal T}^c_l(\epsilon_s) &=& 1-{\mathcal R}^c_l(\epsilon_s) \nonumber\\
	&=& \frac{2M_{\epsilon_s l}^2[1-\cos(2\pi\nu)]}
	{1-2M_{\epsilon_s l}^2\cos(2\pi\nu)+M_{\epsilon_s l}^4} \;.
\label{eq:tp4}	
\end{eqnarray}
\begin{figure}
\includegraphics[width=\columnwidth]{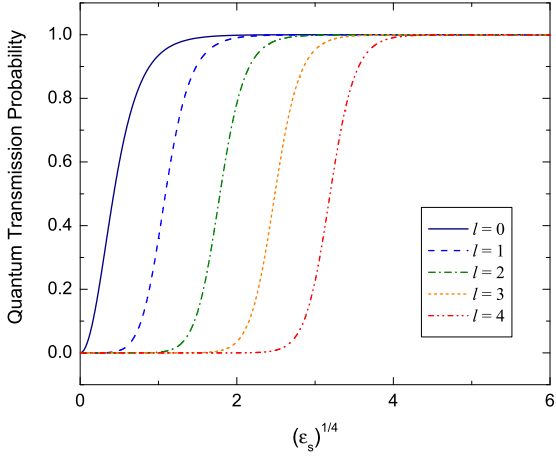}
\caption{(Color online) The quantum transmission (tunneling) probability,
${\mathcal T}^c_l(\epsilon_s)$, through a $-1/r^4$ potential for different
partial waves.
\label{fig:Tc4}}
\end{figure}
It is illustrated in Fig.~\ref{fig:Tc4} for the
first few partial waves. 

The quantum reflection probability,
which also serves as a quantum order parameter for $\epsilon>0$,
is illustrated in Fig.~\ref{fig:Qc4} together with the
quantum order parameter ${\cal Q}_c$ for $\epsilon_s<0$  \cite{gao08a},
to be discussed further in Sec.~\ref{sec:qorder}.
Together, they specify a range of energies, 
where they differ substantially from zero, as the region
in which the quantum effects are important \cite{gao08a}.
In the semiclassical region defined by ${\mathcal R}^c_l\approx 0$,
the effect of the long-range potential on scattering
is fully characterized by the long-range phase shift $\delta^{c}_l$.
In the quantum region, even a single channel scattering has
contributions from multiple paths, which interfere 
with each other \cite{gao08a} to give rise to phenomena such as
the shape resonance. A complete characterization
of the effects of the long-range interaction on scattering
in the quantum regime require all three QDT functions \cite{gao08a}.

From again the $Z^c$ matrix elements, the long-range (transmission)
phase shift $\delta^{c}_l$ and the reflection phase shift
$\phi^{c}_l$, are determined, to within a $2\pi$, by 
\begin{eqnarray}
\sin\delta^{c}_l &=& \frac{Z^c_{gs}-Z^c_{fc}}
	{\sqrt{(Z^c_{fs}+Z^c_{gc})^2+(Z^c_{fc}-Z^c_{gs})^2}} \;,\\
	&=&\frac{\cos(\pi\nu-\frac{1}{2}\pi\nu_0)
		+(-1)^l M_{\epsilon_s l}^2\sin(\pi\nu-\frac{1}{2}\pi\nu_0)}
		{\sqrt{1-2M_{\epsilon_s l}^2\cos(2\pi\nu)+M_{\epsilon_s l}^4}} \;,\\
\cos\delta^{c}_l &=& \frac{Z^c_{fs}+Z^c_{gc}}
	{\sqrt{(Z^c_{fs}+Z^c_{gc})^2+(Z^c_{fc}-Z^c_{gs})^2}} \;,\\
	&=&\frac{\sin(\pi\nu-\frac{1}{2}\pi\nu_0)
		+(-1)^l M_{\epsilon_s l}^2\cos(\pi\nu-\frac{1}{2}\pi\nu_0)}
		{\sqrt{1-2M_{\epsilon_s l}^2\cos(2\pi\nu)+M_{\epsilon_s l}^4}} \;,
\end{eqnarray}
and
\begin{eqnarray}
\sin\phi^{c}_l &=& \frac{Z^c_{fc}+Z^c_{gs}}
	{\sqrt{(Z^c_{fs}-Z^c_{gc})^2+(Z^c_{fc}+Z^c_{gs})^2}} \;,\\
	&=& \cos(\pi\nu_0/2) \;,
\label{eq:sinphi} \\
\cos\phi^{c}_l &=& \frac{Z^c_{gc}-Z^c_{fs}}
	{\sqrt{(Z^c_{fs}-Z^c_{gc})^2+(Z^c_{fc}+Z^c_{gs})^2}} \;,\\
	&=& -\sin(\pi\nu_0/2) \;,
\label{eq:cosphi}	
\end{eqnarray}
from which we obtain $\phi^{c}_l = l\pi/2+3\pi/4$, independent of energy.
This value for $\phi^{c}_l$ implies that, for the $-1/r^4$ potential,
the two reflection amplitudes are related by $r^{(io)}_l=i r^{(oi)}_l$
(where $i=\sqrt{-1}$).

\subsection{Quantum order parameter and other QDT functions below the threshold}
\label{sec:qorder}

\begin{figure}
\includegraphics[width=\columnwidth]{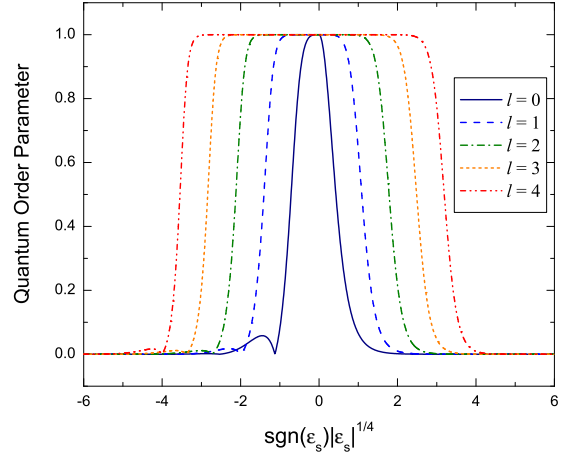}
\caption{(Color online) The quantum order parameter for the $-1/r^4$ potential, 
as represented by the quantum reflection probability $\mathcal{R}^c_l$ for positive energies and 
by the quantum order parameter $\mathcal{Q}^c_l$ for negative energies \cite{gao08a}.
The region where the quantum order parameter is approximately zero
corresponds to the semiclassical region. The region where
it differ substantially from zero corresponds to the quantum region \cite{gao08a}.
\label{fig:Qc4}}
\end{figure}

As discussed in Ref.~\cite{gao08a}, the QDT for negative energies
can be formulated using either the $W^c$ matrix, or three
QDT functions including two phases $\Phi^c_l$ and $\Theta^c_l$, 
and one amplitude $D^c_l$. The latter formulation is convenient for, 
e.g., understanding the semiclassical limit away from the threshold.

Following Ref.~\cite{gao08a}, the functions $\Phi^c_l$, $\Theta^c_l$, 
and $D^c_l$ for $n=4$ are obtained in a straightforward manner 
from the $W^c$ matrix of Sec.~\ref{sec:qdtfuncs}. We have 
\begin{eqnarray}
\sin\Phi^c_l &=& -W^c_{f-}/\sqrt{(W^c_{f-})^2+(W^c_{g-})^2} \;,\\
	&=& -\sin(\pi\nu/2)\frac{1+M_{\epsilon_s l}^2}
		{\sqrt{1-2M_{\epsilon_s l}^2\cos(\pi\nu)+M_{\epsilon_s l}^4}} \;,
\end{eqnarray}
\begin{eqnarray}
\cos\Phi^c_l &=& W^c_{g-}/\sqrt{(W^c_{f-})^2+(W^c_{g-})^2} \;,\\
	&=& \cos(\pi\nu/2)\frac{1-M_{\epsilon_s l}^2}
		{\sqrt{1-2M_{\epsilon_s l}^2\cos(\pi\nu)+M_{\epsilon_s l}^4}} \;.
\end{eqnarray}
Together, they determine the quantum phase $\Phi^c_l$ to within a $2\pi$.
The amplitude $D^c_l$ is given by
\begin{eqnarray}
D^c_l &=& \sqrt{(W^c_{f+})^2+(W^c_{g+})^2} \;,\\
	&=& \frac{1}{2^{1/2}M_{\epsilon_s l}}
		\sqrt{1+2M_{\epsilon_s l}^2\cos(\pi\nu)+M_{\epsilon_s l}^4} \;,
\end{eqnarray}
and the phase $\Theta^c_l$ is determined by
\begin{eqnarray}
\sin\Theta^c_l &=& -W^c_{f+}/D^c_l \;,\\
	&=& \sin(\pi\nu/2)\frac{1-M_{\epsilon_s l}^2}
		{\sqrt{1+2M_{\epsilon_s l}^2\cos(\pi\nu)+M_{\epsilon_s l}^4}} \;,
\end{eqnarray}
\begin{eqnarray}
\cos\Theta^c_l &=& W^c_{g+}/D^c_l \;,\\
	&=& -\cos(\pi\nu/2)\frac{1+M_{\epsilon_s l}^2}
		{\sqrt{1+2M_{\epsilon_s l}^2\cos(\pi\nu)+M_{\epsilon_s l}^4}} \;,
\end{eqnarray}
which give $\Theta^c_l$ to within a $2\pi$.

For most conventional applications,
the most important QDT function for negative energies is the
quantum phase $\Phi^c_l$, or the closely related $\chi^c_l$ 
function defined by $\chi^c_l=W^{c}_{f-}/W^{c}_{g-}=-\tan\Phi^c_l$ \cite{gao08a}.
They, together with the quantum defect, 
determine the bound spectrum, to be discussed 
further in Sec.~\ref{sec:boundsp}. In a multichannel formulation \cite{gao05a,gao11b},
the same functions, and a short-range $K^c$ matrix, 
characterize not only the bound spectrum, but
also the Fano-Feshbach resonances.
A complete understanding of the negative energy states \cite{gao08a}
will, however, generally require all three QDT functions,
namely $D^c_l$ and $\Theta^c_l$, in addition to $\Phi^c_l$. 
Scattering at negative energy, to be discussed in Sec.~\ref{sec:scatne},
is one such example.

Another useful QDT function for negative energies is the 
quantum ``order parameter'', ${\mathcal Q}^c_l$, introduced in
Ref.~\cite{gao08a}. From the $W^c$ matrix, we obtain
for $-1/r^4$ potential
\begin{eqnarray}
{\mathcal Q}^c_l &=& -\frac{W^c_{f+}W^c_{f-}+W^c_{g+}W^c_{g-}}
		{\sqrt{[(W^c_{f+})^2+(W^c_{g+})^2][(W^c_{f-})^2+(W^c_{g-})^2]}} \;,\nonumber\\
	&=& -\cos(\Phi^c_l-\Theta^c_l) \;,\\
	&=& \frac{1-M_{\epsilon_s l}^4}
		{\sqrt{1-2M_{\epsilon_s l}^4\cos(2\pi\nu)+M_{\epsilon_s l}^8}}\;.
\end{eqnarray}
In terms of ${\mathcal Q}^c_l$, the quantum and the semiclassical regions
of energies below the threshold can be characterized by
${\mathcal Q}^c_l\neq 0$ and ${\mathcal Q}^c_l\approx 0$, respectively.
It is only in the semiclassical region of ${\mathcal Q}^c_l\approx 0$
that the semiclassical description of the bound spectrum \cite{ler70}
would apply. 
The ${\mathcal Q}^c_l$ is illustrated in Fig.~\ref{fig:Qc4} for the first few partial
waves, together with the quantum reflection probability which serves
as the quantum order parameter for positive energies \cite{gao08a}.
From Fig.~\ref{fig:Qc4}, it can be recognized that the quantum
regions of energies, for both positive and negative energies, grow
as $\sim l^4$ for higher partial wave states.

\section{QDT for $-1/r^4$ potential}
\label{sec:qdt}

The QDT for a $-1/r^n$ potential describes two-body interactions
with a $-C_n/r^n$ asymptotic potential in terms of (a) a set of universal
QDT functions that are determined by $n$ and $l$, 
such as those for $n=4$ presented in the previous section,
(b) a set of scaling factors, such as $\beta_n$ and $s_E$, that are
determined by $C_n$ and the reduced mass, 
and (c) a dimensionless short-range parameter. There are different options for the
short-range parameter, each with its distinctive utilities.
The short-range $K^c$ matrix is
defined by matching the radial wave function $u_{\epsilon l}(r)$,
which is the solution of the radial Schr\"odinger equation with
potential $V(r)$ and satisfies the boundary condition at the origin, 
to a linear combination of the QDT base pair, $f^c_{\epsilon_s l}$ 
and $g^c_{\epsilon_s l}$,
\begin{equation}
u_{\epsilon l}(r) = A_{\epsilon l}[f^c_{\epsilon_s l}(r_s) 
	- K^c(\epsilon,l) g^c_{\epsilon_s l}(r_s)]\;,
\label{eq:wfn}
\end{equation}
at any radius $r>r_0$ where $V(r)$ has become well represented
by its asymptotic behavior of $-C_n/r^n$.
The $K^c(\epsilon,l)$ parameter encapsulates all effects of 
the short-range interaction on the wave function beyond $r_0$. 
It is a short-range $K$ matrix that is 
well defined at all energies and is a meromorphic function of both $\epsilon$
and $l$ \cite{gao08a}. Closely related to the $K^c$ parameter is
a short-range phase $\delta^\mathrm{sr}$ defined by $K^c=\tan\delta^\mathrm{sr}$.

Instead of the $K^c(\epsilon,l)$ parameter, the short-range
physics can also be characterized by related parameters
such as the quantum defect $\mu^c(\epsilon,l)$ or the 
$K^{c0}_l(\epsilon)$ parameter. For any $-1/r^n$ ($n>2$) potential,
the quantum defect $\mu^c$, defined to have a 
range of $0\le \mu^c<1$, is related to the $K^c$ parameter by
$
K^c(\epsilon,l) = \tan[\pi\mu^c(\epsilon,l)+\pi b/2]
$
where $b=1/(n-2)$ \cite{gao08a}.
The $K^{c0}_l(\epsilon)$ parameter is defined by \cite{gao08a}
\begin{eqnarray}
K^{c0}_l(\epsilon) &=& \frac{K^c(\epsilon, l)-\tan(\pi\nu_0/2)}
	{1+\tan(\pi\nu_0/2)K^c(\epsilon, l)} \;,
\label{eq:Kc0l} \\
	&=&\tan[\pi\mu^c(\epsilon, l)-\pi lb]\;.
\end{eqnarray}
Specializing to the case of $n=4$, they imply the following
relations among the three parameters
\begin{equation}
K^{c0}_l(\epsilon) = \frac{K^c(\epsilon, l)-(-1)^l}
	{1+(-1)^lK^c(\epsilon, l)} \;,
\label{eq:Kc0Kc}	
\end{equation}
and
\begin{equation}
K^{c0}_l(\epsilon) = \left\{
	\begin{array}{ll}
	\tan[\pi\mu^c(\epsilon,l)] \;, & l=\text{even} \\
	-\cot[\pi\mu^c(\epsilon,l)] \;,& l=\text{odd}
	\end{array} \right.\;,
\label{eq:Kc0muc}	
\end{equation}
for the $-1/r^4$ potential. All parameters contains the same amount
of physics and are well defined at all energies.
Their different utilities \cite{gao08a,gao09a}
are further illustrated in subsequent examples and discussions.

\subsection{Scattering}

\subsubsection{Scattering at positive energies}
\label{sec:scatpe}

The single-channel scattering is described in QDT by \cite{gao08a}
\begin{equation}
K_l \equiv \tan\delta_l = 
	\left(Z^{c(n)}_{gc}K^c -Z^{c(n)}_{fc}\right)
	\left(Z^{c(n)}_{fs}-Z^{c(n)}_{gs}K^c \right)^{-1} \;,
\label{eq:qdtpe}
\end{equation}
which gives the scattering phase shift.
Here $Z^{c(n)}_{xy}(\epsilon_s,l)$ for $n=4$ are elements
of the $Z^c$ matrix given in the previous section.
Implied in the QDT description is the physics that
a long-range interaction of the type of $-1/r^n$ with $n>2$ 
affects the scattering not only through a long-range phase shift, as is the case
for $n<2$, but also through quantum reflection and tunneling.
This physics is more transparently reflected in an equivalent QDT description
using reflection and transmission amplitudes \cite{gao08a}. 
For example, Eq.~(\ref{eq:qdtpe}) can be written as \cite{gao08a}
\begin{equation}
K_l = \frac{\sin(\delta^\mathrm{sr}+\delta^{c(n)}_l)+\sqrt{{\mathcal R}^{c(n)}_l}\sin(\delta^\mathrm{sr}-\phi^{c(n)}_l)}
	{\cos(\delta^\mathrm{sr}+\delta^{c(n)}_l)-\sqrt{{\mathcal R}^{c(n)}_l}\cos(\delta^\mathrm{sr}-\phi^{c(n)}_l)} \;,
\label{eq:qdtpeR}
\end{equation}
where $\delta^{c(n)}_l$, ${\mathcal R}^{c(n)}_l$, and $\phi^{c(n)}_l$ for $n=4$ are
QDT functions given in Sec.~\ref{sec:qdtfuncs}.
It is only for sufficiently high energies where ${\mathcal R}^{c(n)}_l\approx 0$
that the effects of the long-range potential reduce to that of a long-range
phase shift, as represented by 
\begin{equation}
K_l \stackrel{{\mathcal R}^{c(n)}_l\rightarrow 0}{\sim} 
	\tan(\delta^\mathrm{sr}+\delta^{c(n)}_l) \;.
\end{equation}
In the quantum region of ${\mathcal R}^{c(n)}_l\neq 0$, even a single
channel scattering has contributions from multiple paths \cite{gao08a}.
It is the interference among such contributions that gives rise to
the shape resonance structures. No such structure exists in the
either the reflection or the transmission probabilities themselves
(see Figs.~\ref{fig:Tc4} and \ref{fig:Qc4}).

Both QDT formulations of scattering, Eqs.~(\ref{eq:qdtpe}) and (\ref{eq:qdtpeR}),
are completely general, and applicable 
regardless of whether or how the short-range parameters may depend on energy and/or $l$.
For $n=4$, it is clear from the discussion of ${\mathcal R}^{c(n)}_l$ 
in Sec.~\ref{sec:qdtfuncs} that
the range of energies over which quantum reflection and tunneling remain important
grows with $l$ as $\sim l^4$. A more quantitative characterization of
this range is given by the critical scaled energy $B_c(l)$ \cite{gao10a}, 
to be discussed further in later sections.

\begin{figure}
\includegraphics[width=\columnwidth]{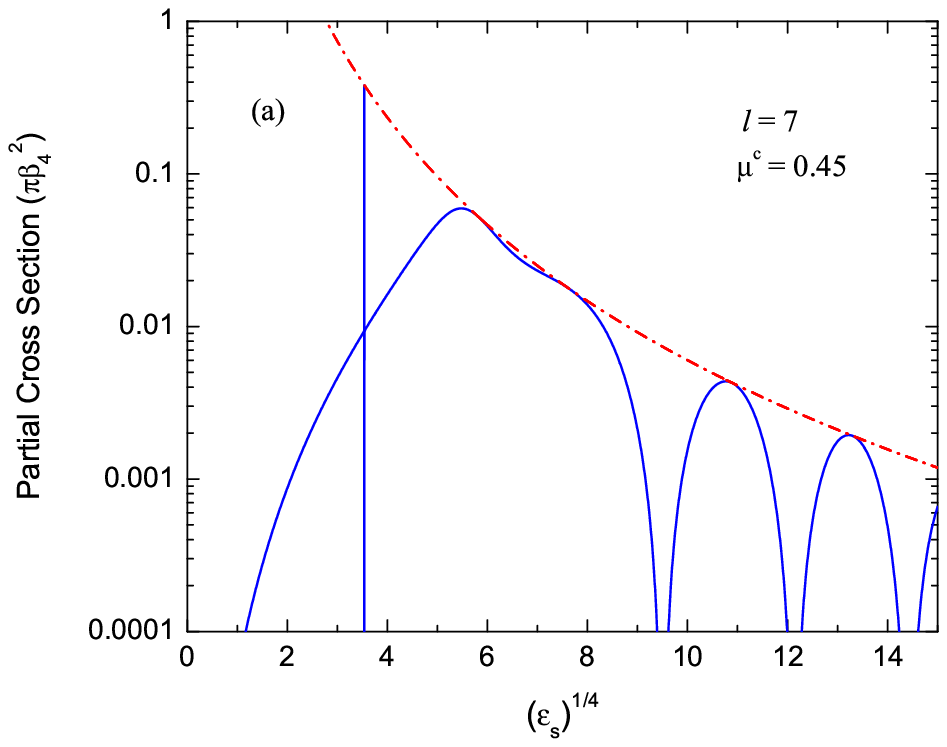}
\includegraphics[width=\columnwidth]{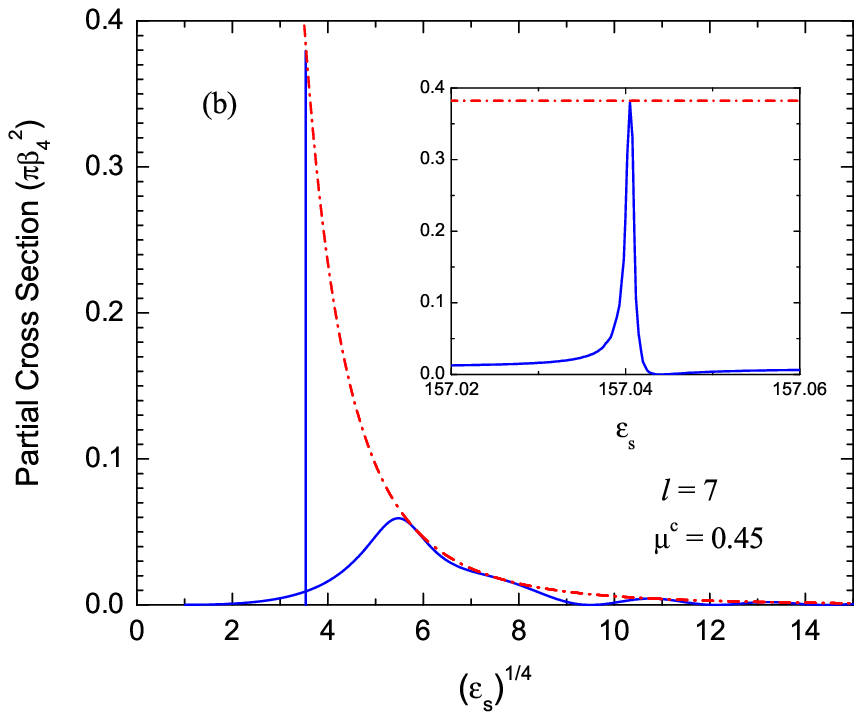}
\includegraphics[width=\columnwidth]{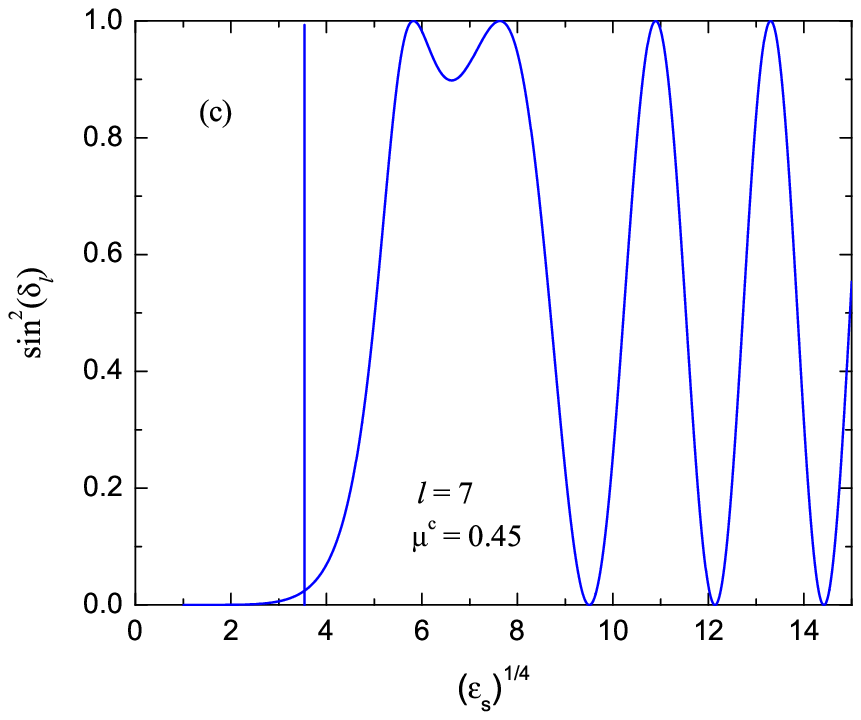}
\caption{(Color online) Scattering characteristics for partial wave $l=7$ 
with a quantum defect of $\mu^c = 0.45$. Both (a) and (b)
show the partial cross section (solid line), with the dash-dot 
line representing the unitarity limit (see text) for $l=7$.
Figure (a) shows the cross section on a LOG scale.
Figure (b) shows it on a linear scale with a close-up of the
narrow feature around $(\epsilon_s)^{1/4}\approx 3.54$.
Figure (c) shows the corresponding $\sin^2\delta_l$.
The set of energies at which the cross section touches the
unitarity limit, or $\sin^2\delta_l$ reaches 1, define
the resonance spectrum to be discussed in Sec.~\ref{sec:rspec}.
\label{fig:pxs4l7}}
\end{figure}
Figure~\ref{fig:pxs4l7} illustrates some of the scattering characteristics
for $n=4$ in partial wave $l=7$.
It assumes an energy-independent quantum defect of 
$\mu^c(\epsilon,l=7) = 0.45$, corresponding to an energy-independent 
$K^c(\epsilon,l=7)\approx -1.37638$ and a $K^{c0}_{l=7}\approx -0.158384$. 
It is an example used here to motivate the concept of the resonance 
spectrum and to illustrate the existence of multiple shape resonances
for sufficiently large $l$. Both were discussed briefly in Ref.~\cite{gao10a},
and will be discussed in more detail in later sections.

Figure~\ref{fig:pxs4l7}(a) shows, on a LOG scale, the partial scattering cross section
\[
\sigma_l/(\pi\beta_4^2) = [4(2l+1)/\epsilon_s]\sin^2\delta_l \;,
\]
for $l=7$, over a range of energies of $0<\epsilon_s<50625$ or $0<\epsilon_s^{1/4}<15$. 
Figure~\ref{fig:pxs4l7}(b) shows the same cross section on a linear
scale, with a closer look at the narrow structure
around $\epsilon_s^{1/4}\approx 3.54$ ($\epsilon_s\approx 157$).
Figure~\ref{fig:pxs4l7}(c) shows, instead of the partial cross section, the
corresponding $\sin^2\delta_l=|S_l-1|^2/4$, where $S_l=e^{i2\delta_l}$ is
the single-channel $S$ matrix \cite{gao08a,tay06}. 
The $\sin^2\delta_l$ is basically the partial
cross section scaled by its unitarity limit, given by 
$\sigma^{\mathrm{max}}_l/(\pi\beta_4^2)=4(2l+1)/\epsilon_s=60/\epsilon_s$
for $l=7$.
Together, they show that there are considerable structures in scattering.
The structures at higher energies are less prominent in the cross section, 
but only because of the constraint of the unitarity limit.
With proper scaling, of both energy and the cross section, there is little
difference among the last 4 structures shown in Fig.~\ref{fig:pxs4l7}.

Without the concepts of resonance spectrum, width function, and diffraction 
resonance \cite{gao10a}, the structures shown in Fig.~\ref{fig:pxs4l7}
are easily missed, or unexplained.
Potential existence of narrow resonances, such as the first one
in Fig.~\ref{fig:pxs4l7}, is a general characteristic of 
low-energy heavy particle (anything other than the electron) 
neutral-neutral and charge-neutral scattering.
Without the resonance spectrum identifying the existence and the locations 
of such resonances, a standard numerical calculation, 
which is always performed on a discrete energy mesh, 
can easily miss some or all of them. They also occur far below the barrier
where numerical stability becomes a problem.
The concept of the diffraction resonance will help to
distinguish the last three resonances from the first two,
and the width function will help to provide precise characterizations
of all resonances. They will be discussed in Secs.~\ref{sec:qdtres}
and \ref{sec:uwf}.
 
\subsubsection{Scattering at negative energies}
\label{sec:scatne}

\begin{figure}
\includegraphics[width=\columnwidth]{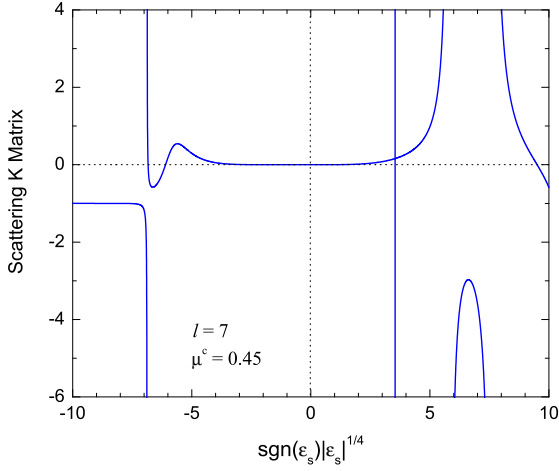}
\caption{(Color online) An illustration of the generalized $K$ matrix, $\widetilde{K}_l$, 
for scattering at negative energies, together with the corresponding
$K_l$ for positive energies, for partial wave $l=7$ and $\mu^c=0.45$.
\label{fig:scatne}}
\end{figure}

In Ref.~\cite{gao08a}, we introduced, for negative energies, 
the generalized $K$ matrix, $\widetilde{K}_l$,
given in QDT formulation by
\begin{widetext}
\begin{align}
\widetilde{K}_l =& \frac{W^{c(n)}_{f-}+(-1)^lW^{c(n)}_{f+}
	-K^c[W^{c(n)}_{g-}+(-1)^lW^{c(n)}_{g+}]}
	{W^{c(n)}_{f-}-(-1)^lW^{c(n)}_{f+}-K^c[W^{c(n)}_{g-}-(-1)^lW^{c(n)}_{g+}]} \;,\\
	=& \frac{\sin(\Phi^{c(n)}_l+\delta^\mathrm{sr})+(-1)^l(D^{c(n)}_l)^2
	\sin(\Phi^{c(n)}_l-\Theta^{c(n)}_l)
	\sin(\Theta^{c(n)}_l+\delta^\mathrm{sr})}
	{\sin(\Phi^{c(n)}_l+\delta^\mathrm{sr})-(-1)^l(D^{c(n)}_l)^2\sin(\Phi^{c(n)}_l-\Theta^{c(n)}_l)
	\sin(\Theta^{c(n)}_l+\delta^\mathrm{sr})} \;.	
\label{eq:Kt} 
\end{align}
\end{widetext}
where $W^{c(n)}_{xy}$, $\Phi^{c(n)}_l$, $D^{c(n)}_l$, and
$\Theta^{c(n)}_l$ for $n=4$ are the QDT functions 
given in Sec.~\ref{sec:qdtfuncs}.

The $\widetilde{K}_l(\epsilon<0)$ is a generalization of 
the $K_l(\epsilon>0) = \tan\delta_l$ to negative energies. 
It is well defined for all negative energies and give a more complete
characterization of the negative energy states than merely the bound
spectrum. This quantity, together with the concept of 
resonance spectrum \cite{gao10a}, makes our understanding 
of two-body interactions more symmetric and more complete
for both positive and negative energies. 
The bound spectrum is contained within $\widetilde{K}_l(\epsilon<0)$
as the solutions of $\widetilde{K}_l=-1$. The resonance spectrum
is contained within $K_l(\epsilon>0)$, as solutions of $K_l(\epsilon>0) = \infty$.
The generalized $K$ matrix has applications in interaction 
in reduced dimensions \cite{che07},
and in few-body \cite{kha06} and many-body physics \cite{gao04a,gao05b}.
 
In Fig.~\ref{fig:scatne}, we illustrate the $\widetilde{K}_l(\epsilon<0)$,
together with $K_l(\epsilon>0)$, for the case of $l=7$ and $\mu^c = 0.45$.
The positive energy part is the $K_l$ corresponding to the scattering
properties illustrated in Fig.~\ref{fig:pxs4l7}.
Figure~\ref{fig:scatne} also serves to show the general feature that $\widetilde{K}_l$
evolves continuously to $K_l$ at zero energy.
This evolution is however not analytic at $\epsilon=0$, 
with different functional representations in $\epsilon<0$ and
$\epsilon>0$, respectively.

\subsection{Spectrum}
\label{sec:spec}

\subsubsection{Bound spectrum}
\label{sec:boundsp}

In single-channel QDT,
the bound spectrum of a two-body system with 
$-1/r^n$ ($n>2$) type of long-range potential is given rigorously
by the solutions of \cite{gao01,gao08a}
\begin{equation}
\chi^{c(n)}_l(\epsilon_s) = K^c(\epsilon,l) \;.
\label{eq:ubsp}
\end{equation}
Here $\chi^{c(n)}_l=W^{c(n)}_{f-}/W^{c(n)}_{g-}=-\tan\Phi^{c(n)}_l$ 
is a universal function of $\epsilon_s$, uniquely determined 
by $n$ and $l$.
A conceptually useful equivalent of Eq.~(\ref{eq:ubsp}) is
$\Phi^{c(n)}+\delta^\mathrm{sr}=j\pi$, where $j$ is an integer. 

For $n=4$, we have, from the $W^c$ matrix of Sec.~\ref{sec:qdtfuncs},
\begin{equation}
\chi^{c(4)}_l = \tan(\pi\nu/2)\frac{1+M_{\epsilon_s l}^2}{1-M_{\epsilon_s l}^2} \;.
\label{eq:chic4}
\end{equation}
The resulting bound spectrum can be represented in a number of different
manner as to be discussed in Sec.~\ref{sec:specreps}.

\subsubsection{Resonance spectrum}
\label{sec:rspec}

In Ref.~\cite{gao10a}, we introduced the concept of resonance
spectrum as a set of energies at which $\sin^2\delta_l=1$, 
namely the energies at which the partial scattering cross 
section reaches its unitarity limit, as illustrated in
Fig.~\ref{fig:pxs4l7}.
Such locations 
can be determined as the roots of the denominator in
Eq.~(\ref{eq:qdtpe}).
Defining a generalized $\chi^{c(n)}_l$ function
for positive energies as
$
\widetilde{\chi}^{c(n)}_l(\epsilon_s)\equiv Z^{c(n)}_{fs}/Z^{c(n)}_{gs}
$,
the resonance positions can be formulated in a manner similar
to the bound spectrum, as the solutions of
\begin{equation}
\widetilde{\chi}^{c(n)}_l(\epsilon_s) = K^c(\epsilon,l) \;.
\label{eq:ursp}
\end{equation}
For $n=4$, we have from the $Z^c$ matrix of Sec.~\ref{sec:qdtfuncs},
\begin{equation}
\widetilde{\chi}^{c(4)}_l = \tan(\pi\nu/2)
	\frac{1-(-1)^lM_{\epsilon_s l}^2\tan[\pi(\nu-\nu_0)/2]}
	{1+(-1)^lM_{\epsilon_s l}^2\tan[\pi(\nu-\nu_0)/2]} \;.
\label{eq:gchic4}
\end{equation}
The $\widetilde{\chi}^{c(n)}_l$ function can be regarded as an
extension of the $\chi^{c(n)}_l$ function to
positive energies. They evolve continuously, but not analytically, 
into each other across $\epsilon_s=0$, with
\[\lim_{\epsilon_s\rightarrow 0-}\chi^{c(4)}_l
=\lim_{\epsilon_s\rightarrow 0+}\widetilde{\chi}^{c(4)}_l
= \tan(\pi\nu_0/2)=(-1)^l \;.
\]
The $\widetilde{\chi}^{c(n)}_l$ function can be further used to define a phase
$\widetilde{\Phi}^{c(n)}_l$, by $\widetilde{\chi}^{c(n)}_l=-\tan\widetilde{\Phi}^{c(n)}_l$,
as an extension of the quantum phase $\Phi^{c(n)}_l$ to positive energies.
In terms of $\widetilde{\Phi}^{c(n)}_l$, a conceptually useful equivalent 
of Eq.~(\ref{eq:ursp}) is
$\widetilde{\Phi}^{c(n)}+\delta^\mathrm{sr}=j\pi$, where $j$ is an integer.
It is again a natural extension of the bound spectrum.

Similar to a bound spectrum, which
describes, over a set of discrete energies, the rise of a 
phase from zero to a finite value
at the threshold, a resonance spectrum describes its 
subsequent evolution (eventually) back towards zero. It also
describes the evolution of a bound state into continuum, and the evolution of
a resonance into a bound state.
The potential existence of extremely narrow shape resonances for long-range interactions
with $n>2$, as illustrated in Fig.~\ref{fig:pxs4l7}
is another motivation for the introduction of resonance spectrum.
The mathematical and practical necessity for such a concept
will be discussed further in later sections.

\subsubsection{Representations of the spectra}
\label{sec:specreps}
\begin{figure}
\includegraphics[width=\columnwidth]{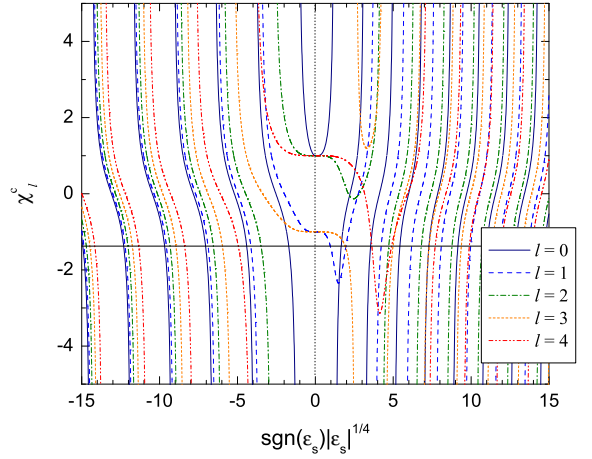}
\caption{(Color online) One representation of the  
spectra for two-body quantum systems with 
$-1/r^4$ type of interactions, including both the 
bound spectra for $\epsilon_s<0$, where the curves plotted
are the $\chi^{c(4)}_l$, and the resonance spectra for
$\epsilon_s>0$, where the curves plotted are the $\widetilde{\chi}^{c(4)}_l$. 
For any two-body quantum system with 
$-1/r^4$ type of long-range potential, the bound spectra
and the resonance spectra are given by the cross points between
the curves plotted and a set of curves representing
$K^c(\epsilon,l)$ for different partial waves $l$. For systems such as ion-atom,
$K^c(\epsilon,l)$ is approximately
an energy- and partial-wave-independent constant, allowing the determination of
the entire rovibrational spectra, and the resonance spectra,
from a single parameter. The curves representing $K^c(\epsilon,l)$ for different $l$
all appear in this case as a single horizontal line,  as illustrated in
the figure. The $\widetilde{\chi}^{c(4)}_l$ functions evolves from
a piecewise monotonically decreasing function of energy to
a piecewise monotonically increasing function of energy at the
critical scaled energy $B_c(l)$.
\label{fig:usp4chic}}
\end{figure}
\begin{figure}
\includegraphics[width=\columnwidth]{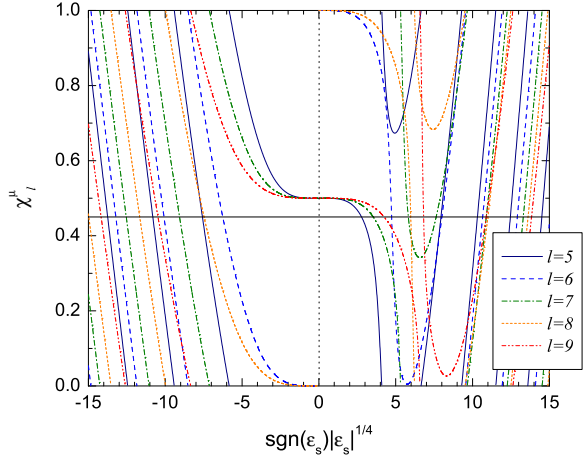}
\caption{(Color online) Representation of the  
spectra as the cross points between
the curves plotted, $\chi^{\mu(4)}_l(\epsilon_s)$ for $\epsilon_s<0$
and $\widetilde{\chi}^{\mu(4)}_l(\epsilon_s)$ for $\epsilon_s>0$, 
and a set of curves representing
$\mu^c(\epsilon,l)$ for different partial waves $l$. 
For ion-atom interaction, $\mu^c(\epsilon,l)$ is approximately
an energy- and partial-wave-independent constant, allowing the determination of
the entire spectra from a single parameter, as illustrated by a 
horizontal line in
the figure corresponding to $\mu^c=0.45$. 
\label{fig:usp4muc}}
\end{figure}
There are a number of different representations of both the bound and
the resonance spectra, corresponding to descriptions of
the short-range physics using different parameters such as
$K^c(\epsilon,l)$, $\mu^c(\epsilon,l)$, or $K^{c0}_l(\epsilon)$. 
They have different utilities and applications,
and offer different physical insights.

\textbf{$K^c$ representation}:
Figure~\ref{fig:usp4chic} gives the most direct representation of 
Eqs.~(\ref{eq:ubsp}) and (\ref{eq:ursp}). It represents the bound
spectra as the crossing points between the $K^c(\epsilon,l)$ and
the $\chi^{c(n)}_l$ function for $\epsilon<0$, and the resonance spectra
as the crossing points between the $K^c(\epsilon,l)$ and 
$\widetilde{\chi}^{c(n)}_l$ function for $\epsilon>0$.
It is the base representation that is the most convenient 
for most computational purposes.

For negative energies, $\chi^{c(n)}_l$ is a piecewise monotonically
decreasing function of energy with $d\widetilde{\chi}^{c(n)}_l/d\epsilon_s<0$. 
It evolves into $\widetilde{\chi}^{c(n)}_l$
at zero energy. For positive energies,
$\widetilde{\chi}^{c(n)}_l$ continues to be piecewise monotonically
decreasing until it reaches the critically scaled
energy $B_{c}(l)$ defined by 
$\left.d\widetilde{\chi}^{c(n)}_l/d\epsilon_s\right|_{B_c(l)}=0$.
Above $B_c(l)$, $\widetilde{\chi}^{c(n)}_l$ evolves into a
piecewise monotonically increasing function of energy with
$d\widetilde{\chi}^{c(n)}_l/d\epsilon_s>0$. 
In terms of the closely related quantum phase,
the critical scaled energy $B_c(l)$ corresponds to the scaled 
energy at which the quantum phase evolves from monotonically increasing
to monotonically decreasing with energy.

\textbf{$\mu^c$ representation}:
In the second representation of the spectrum, the 
short-range physics is described using the $\mu^c(\epsilon,l)$ parameter.
Specifically, Eqs.~(\ref{eq:ubsp}) and (\ref{eq:ursp}) 
can be rewritten as
\begin{equation}
\chi^{\mu(n)}_l(\epsilon_s) = \mu^c(\epsilon,l) \;,
\label{eq:bspmuc}
\end{equation}
for the bound spectrum, and
\begin{equation}
\widetilde{\chi}^{\mu(n)}_l(\epsilon_s) = \mu^c(\epsilon,l) \;,
\label{eq:rspmuc}
\end{equation}
for the resonance spectrum.
Here $\chi^\mu_l$ is defined in terms of $\chi^c_l$ as 
$\chi^\mu_l = [\tan^{-1}(\chi^c_l)-\pi b/2]/\pi$, in which
$\tan^{-1}(\chi^c_l)$ is taken to be within a range of $\pi$
of $[\pi b/2, \pi+\pi b/2)$, where $b=1/(n-2)$. The $\widetilde{\chi}^\mu_l$ 
is defined in terms of $\widetilde{\chi}^c_l$ in a similar manner.
In this representation, the spectra are given by the
crossing points between the $\mu^c(\epsilon,l)$ and
the $\chi^{\mu(n)}_l$ function for $\epsilon<0$, and 
$\widetilde{\chi}^{\mu(n)}_l$ function for $\epsilon>0$,
as illustrated in Fig.~\ref{fig:usp4muc} for $l=5$-$9$.
This representation is convenient for the visualization of
the semiclassical limit \cite{ler70,fla99,Friedrich2004} and for understanding the structure
of the rovibrational states around the threshold and the corresponding
classification of molecules (molecular ions to be more precise here) \cite{gao04b}.
Similar to $\widetilde{\chi}^{c(n)}_l$, the $\widetilde{\chi}^{\mu(n)}_l$ function
evolves from being monotonically decreasing function of energy
to being monotonically increasing function of energy at $B_c(l)$.

The semiclassical limit corresponds to regions in Fig.~\ref{fig:usp4muc}
where $\chi^\mu_l$ (or $\widetilde{\chi}^\mu_l$) becomes a set of
equally-spaced parallel straight lines versus $(-\epsilon_s)^{1/4}$ for
$\epsilon<0$ (versus $\epsilon_s^{1/4}$ for $\epsilon>0$).
The QDT, being an exact quantum theory, thus also provides a
framework for testing various semiclassical approximations \cite{gao99b}
such as the WKB approximation \cite{ler70,fla99,Friedrich2004}.
From Fig.~\ref{fig:usp4muc}, it is clear that
the greater the $l$, the greater the
range of energies around the threshold in which the WKB
approximation fails. This range is characterized by the quantum
order parameter of Sec.~\ref{sec:qdtfuncs}, and grows as $l^4$
both below and above the threshold. In the quantum region of negative
energies, the number of states is reduced compared to what is to be expected from 
the WKB theory \cite{ler70,fla99,Friedrich2004}. They are pushed into the
quantum region above the threshold.

\begin{figure}
\includegraphics[width=\columnwidth]{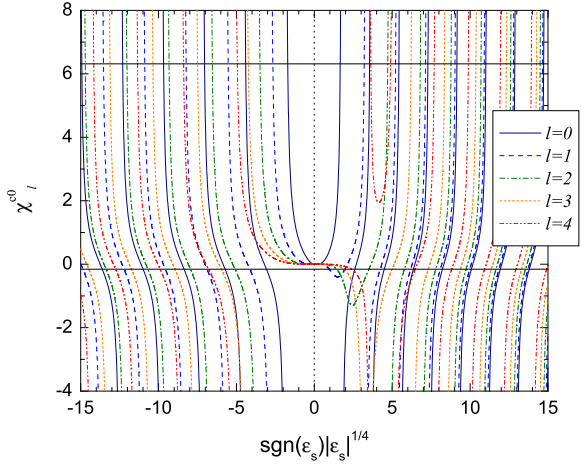}
\caption{(Color online) Representation of the  
spectra as the cross points between
the curves plotted, $\chi^{c0(4)}_l(\epsilon_s)$ for $\epsilon_s<0$
and $\widetilde{\chi}^{c0(4)}_l(\epsilon_s)$ for $\epsilon_s>0$, 
and a set of curves representing
$K^{c0}_l(\epsilon)$ for different partial waves $l$. 
Unlike the $K^c(\epsilon,l)$ and the $\mu^c(\epsilon,l)$ parameters,
$K^{c0}_l(\epsilon)$ is generally different for different $l$.
For cases, such as ion-atom interactions, where $K^c$ and $\mu^c$ are 
approximately energy- and partial-wave-independent, 
there are approximately two \textit{related} energy-independent $K^{c0}_l$s
given by Eq.~(\ref{eq:Kc0Kc}) or (\ref{eq:Kc0muc}), with one 
describing all even partial waves and one describing all odd
partial waves. For the example of $\mu^c=0.45$, $K^{c0}_l\approx 6.31375$
for all even partial waves, and $K^{c0}_l\approx -0.158384$ for all odd
partial waves, as illustrated by two horizontal lines.
\label{fig:usp4Kc0}}
\end{figure}
\textbf{$K^{c0}$ representation}:
The third representation of the spectra corresponds to the description
of the short-range physics using the 
$K^{c0}_l$ parameter.
Eqs.~(\ref{eq:ubsp}) and (\ref{eq:ursp}) can be rewritten as
\begin{equation}
\chi^{c0(n)}_l(\epsilon_s) = K^{c0}_l(\epsilon) \;,
\label{eq:bspKc0}
\end{equation}
for bound spectrum, and
\begin{equation}
\widetilde{\chi}^{c0(n)}_l(\epsilon_s) = K^{c0}_l(\epsilon) \;,
\label{eq:rspKc0}
\end{equation}
for the resonance spectrum. Here
\begin{equation}
\chi^{c0(n)}_l(\epsilon_s) = \frac{\chi^{c(n)}_l(\epsilon_s)-\tan(\pi\nu_0/2)}
	{1+\tan(\pi\nu_0/2) \chi^{c(n)}_l(\epsilon_s)} \;.
\label{eq:chic0}
\end{equation}
\begin{equation}
\widetilde{\chi}^{c0(n)}_l(\epsilon_s) = \frac{\widetilde{\chi}^{c(n)}_l(\epsilon_s)-\tan(\pi\nu_0/2)}
	{1+\tan(\pi\nu_0/2) \widetilde{\chi}^{c(n)}_l(\epsilon_s)} \;.
\label{eq:gchic0}
\end{equation}
For $n=4$, we obtain
\begin{equation}
\chi^{c0(4)}_l = 
	\frac{\tan(\pi\Delta\nu/2)+(-1)^lM_{\epsilon_s l}^2}
	{1+(-1)^lM_{\epsilon_s l}^2\tan(\pi\Delta\nu/2)} \;,
\label{eq:chic04}
\end{equation}
\begin{equation}
\widetilde{\chi}^{c0(4)}_l = \tan(\pi\Delta\nu/2)
	\frac{1-M_{\epsilon_s l}^2}
	{1-M_{\epsilon_s l}^2\tan^2(\pi\Delta\nu/2)} \;.
\label{eq:gchic04}
\end{equation}
Figure~\ref{fig:usp4Kc0} illustrates this representation of
the spectra. It is the most convenient representation for developing QDT expansion 
around the threshold \cite{gao09a}, and for understanding the relationship
between bound-state or resonance positions and the scattering length (see Sec.~\ref{sec:usp4}). 
All three representations of the spectra are general representations since 
$K^c(\epsilon,l)$, $\mu^c(\epsilon,l)$, and $K^{c0}_l(\epsilon)$ are
well defined at all energies and for all partial waves.
Similar to $\widetilde{\chi}^{c(n)}_l$ and $\widetilde{\chi}^{\mu(n)}_l$,
the $\widetilde{\chi}^{c0(n)}_l$ function
evolves from monotonically decreasing below $B_c(l)$
to monotonically increasing above $B_c(l)$.

\begin{table}
\caption{Energy bins for partial waves $l=5$ through $l=9$.
The numbers in the parenthesis represent powers of ten.
The $i$-th bound state of angular momentum $l$, with $i=1$ 
corresponding to the least-bound state, is to be
found within $B_{-i}(l)\le\epsilon_s< B_{-i+1}(l)$ for $i>1$,
and within $B_{-1}(l)\le \epsilon_s< 0$ for $i=1$.
Shape resonances of angular momentum $l$ can only exist
between $0<\epsilon_{s}<B_c(l)$. 
The $i$-th diffraction resonance of angular momentum $l$,
defined as a resonance with negative width, 
is to be found within $B_{i-1}(l)<\epsilon_s\le B_i(l)$ for
$i\ge 1$. A zeroth diffraction resonance may exist 
within $B_c(l)< \epsilon_s\le B_{0}(l)$, depending on the
quantum defect.
\label{tb:ebins}}
\begin{ruledtabular}
\begin{tabular}{lrrrrr}
$B_x$    & $l=5$      & $l=6$      & $l=7$      & $l=8$      & $l=9$   \\
\hline
$B_{4}$  & 7.6023(4)  & 1.0928(5)  & 9.8278(4)  & 1.3698(5)  & 1.8587(5)   \\
$B_{3}$  & 4.5573(4)  & 6.8266(4)  & 5.9353(4)  & 8.5880(4)  & 1.2030(5)   \\
$B_{2}$  & 2.4932(4)  & 3.9430(4)  & 3.2379(4)  & 4.9310(4)	 & 7.2101(4)   \\
$B_{1}$  & 1.1826(4)  & 2.0212(4)  & 1.4762(4)  & 2.4382(4)  & 3.8074(4)  \\
$B_{0}$  & 4.2509(3)  & 8.3146(3)  & 3.8643(3)  & 8.1816(3)  & 1.5005(4)  \\
$B_{c}$  & 5.9473(2)  & 1.0880(3)  & 1.8636(3)  & 3.0457(3)	 & 4.7668(3)  \\
\hline
$B_{-1}$ & -3.6177(3) & -5.2716(3) & -7.3558(3) & -9.9186(3) & -1.3008(4) \\
$B_{-2}$ & -1.4428(4) & -1.9668(4) & -2.5999(4) & -3.3520(4) & -4.2327(4)  \\
$B_{-3}$ & -3.7224(4) & -4.8531(4) & -6.1818(4) & -7.7233(4) & -9.4923(4)  \\
$B_{-4}$ & -7.7955(4) & -9.8375(4) & -1.2189(5) & -1.4869(5) & -1.7899(5)  
\end{tabular}										
\end{ruledtabular}									
\end{table}
There are many applications of these spectra, which are the equivalents
and the generalizations of the atomic Rydberg formula to charge-neutral
quantum systems. They relate bound spectrum to scattering and vice versa \cite{gao98b,gao01}, 
and provide a systematic understanding for both. One such application is
the concept of energy bins \cite{gao00,chi10,gao10a}:
the ranges of energies over which a bound or
a resonance state is to be found. They
have been given for the first few partial waves, $l=0$-$4$,
in Ref.~\cite{gao10a}.
Table~\ref{tb:ebins} gives the bins for higher partial waves $l=5$-$9$.
In all representations of the spectra, they are determined by the set
of scaled energies at which the relevant $\chi_l$ function has evolved back to
its value at the threshold. In terms of the quantum phase due to the 
long-range potential, $\Phi^{c(n)}_l$ and
$\widetilde{\Phi}^{c(n)}_l$, they correspond
to a set of scaled energies at which the quantum phase differs from its value
at the threshold by an integer multiple of $\pi$.
Even in cases with substantial energy variations in the short-range parameter,
the bins still give the number of states due to the long-range potential.

The energy bin concept  \cite{gao00,chi10,gao10a} is useful not only in single-channel but also 
in multichannel formulations \cite{gao05a,gao11b}, where it
can be used, for instance, to estimate the number of Fano-Feshbach resonances.
A detailed example will be given in a separate publication on MQDT
for ion-atom interactions. 
Further applications of the spectra will
be discussed in Secs.~\ref{sec:ub4} and \ref{sec:discuss}.
We point out that for the $s$ wave bound states, 
Raab and Friedrich have developed a quantization rule by extrapolating between
the quantum threshold behavior and the semiclassical behavior \cite{raa09}.

\subsection{QDT description of scattering resonances}
\label{sec:qdtres}

The resonance spectrum gives only resonance positions. 
The QDT equation for scattering, Eq.~(\ref{eq:qdtpe}),
contains additional information on scattering resonances including
their widths and backgrounds that can be further extracted.

Around a scattering resonance located at $\epsilon_{sl}$,
which is one of the solutions of Eq.~(\ref{eq:ursp}), or equivalently
Eq.~(\ref{eq:rspmuc}) or (\ref{eq:rspKc0}),
the scattering $K$ matrix as given by Eq.~(\ref{eq:qdtpe}) 
can be written as
\begin{equation}
K_l(\epsilon)=K_{\text{bg}l}(\epsilon)-
	\frac{1}{2}\frac{\Gamma_{sl}}{\epsilon_s-\epsilon_{sl}} \;,
\label{eq:resK}	
\end{equation}
where both the background term, $K_{\text{bg}l}(\epsilon)$,
and the scaled width $\Gamma_{sl}$ can be given in terms
of a single function $f_\Gamma$ defined by 
\begin{equation}
f_\Gamma \equiv \frac{\left(Z^{c(n)}_{gc}K^c -Z^{c(n)}_{fc}\right)/Z^{c(n)}_{gs}}
		{\left[\left(\widetilde{\chi}^{c(n)}_l- K^c\right)/(\epsilon_s-\epsilon_{sl})\right]} \;.
\end{equation}
Specifically,
\begin{align}
K_{\text{bg}l}(\epsilon) &= \frac{f_\Gamma(\epsilon_s)-f_\Gamma(\epsilon_{sl})}
	{\epsilon_s-\epsilon_{sl}} \;,\\
	\Gamma_{sl} &=-2f_\Gamma(\epsilon_{sl}) \;.
\end{align}
The function $f_\Gamma$ is regular at $\epsilon_{sl}$,
with a value of $f_\Gamma(\epsilon_{sl})=\lim_{\epsilon_s\rightarrow\epsilon_{sl}}f_\Gamma(\epsilon_{s})$.
Using the property of $\det(Z^{c(n)})=1$ \cite{gao08a}, we obtain
\begin{equation}
\Gamma_{sl} = -2\left\{\left[Z^{c(n)}_{gs}(\epsilon_{sl},l)\right]^2
	\left.\left[\frac{d\widetilde{\chi}^{c(n)}_l}{d\epsilon_s}
	-\frac{dK^c}{d\epsilon_s}\right]\right|_{\epsilon_{sl}}\right\}^{-1} \;.
\label{eq:Gammal}
\end{equation}	
For most true single-channel problems, 
the energy dependence of the $K^c$ is negligible,
and $\Gamma_{sl}$ reduces to a universal function of the scaled resonance
position, given by
\begin{equation}
\gamma^{(n)}_{sl}(\epsilon_{sl}) = -2\left\{\left.\frac{d\widetilde{\chi}^{c(n)}_l}{d\epsilon_s}
	\right|_{\epsilon_{sl}}\left[Z^{c(n)}_{gs}(\epsilon_{sl},l)\right]^2\right\}^{-1} \;.
\label{eq:gammal}
\end{equation}
It will be called the universal width function. It is a function of the scaled
resonance position that is uniquely determined by $n$ and $l$, 
and is given in terms of the QDT functions defined earlier.

This QDT description of scattering resonance is generally applicable to
any long-range potential of the form of $-C_n/r^n$ with $n>2$.
The $\gamma^{(n)}_{sl}$ for $n=4$ is discussed further 
in the next section.
The more general expression for the scaled width,
Eq.~(\ref{eq:Gammal}), will be useful in cases where the energy
dependence of $K^c$ is not negligible. These include some cases
of electron-atom interactions, and maybe more importantly,
some cases corresponding to effective single-channel descriptions 
of Fano-Feshbach resonances, for which the effective $K^c$ parameter
can have substantial energy dependence around a narrow resonance \cite{gao11b}.

\section{Single-channel universal behaviors for $-1/r^4$ potential}
\label{sec:ub4}

Embedded in the QDT descriptions of spectra and scattering resonances
are a set of universal properties followed in varying degrees by
virtually all single-channel charge-neutral systems in a range of energies
around the threshold. They correspond to a set of conclusions that can
be drawn from the QDT formulation under the assumption that the
short-range parameter, $K^c(\epsilon,l)$, $\mu^c(\epsilon,l)$, or $K^{c0}_l(\epsilon)$,
is independent of energy.
Mathematically, they can also be defined rigorously as the universal
property at length scale $\beta_4$, emerging in the limit of
other length scales going to zero in comparison \cite{gao04a,gao05b,kha06}.
Among these properties, there is a subset of conclusions that
can be drawn under the further assumption that the parameter 
$K^c(\epsilon,l)$ or $\mu^c(\epsilon,l)$ [but not $K^{c0}_l(\epsilon)$] 
are not only independent of energy, but also independent of $l$.
This subset is applicable, e.g., to ion-atom interactions, 
for which they imply a set of relations among interactions
in different partial waves \cite{gao01,gao00,gao04b,LG12}. 
They are not applicable to electron- or positron-atom interactions,
for which the relationship between interactions in different
$l$ depends on the details of the short-range potential.

\subsection{Universal width function}
\label{sec:uwf}
\begin{figure}
\includegraphics[width=\columnwidth]{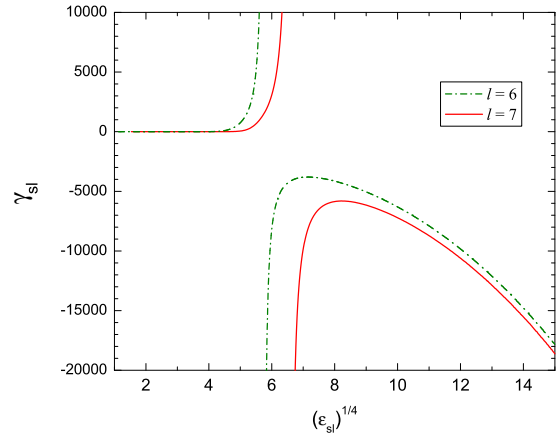}
\caption{(Color online) Illustrations of the universal width function
for $l=6$ and 7. The width diverges at the critical scaled energy $\epsilon_{sl}=B_c(l)$.
It is positive below $B_c(l)$, in the region corresponding to shape resonances,
and negative above $B_c(l)$, in the region corresponding to diffraction resonances.
\label{fig:gamma4}}
\end{figure}

We begin our discussion of universal behaviors with the universal width function,
as it is required for further understanding and interpretation of the resonance spectra.
Under the assumption of the short-range parameter being independent of
energy, the scaled width $\Gamma_{sl}$ of a scattering resonance, given
generally by Eq.~(\ref{eq:Gammal}), reduces to the universal width function 
$\gamma_{sl}$ given by
Eq.~(\ref{eq:gammal}). It implies that while the position of a scattering resonance
depends generally on the short-range parameter such as $K^c$,
the width of the resonance, as a function of the scaled resonance
position $\epsilon_{sl}$, follows a universal behavior for an energy-independent $K^c$.

The most important characteristic of the universal width function is
that it changes sign and diverges at a critical scaled energy 
$B_{c}(l)$. 
Below $B_c(l)$, $\widetilde{\chi}^{c(n)}_l$ is a
piecewise monotonically decreasing function of energy with
$d\widetilde{\chi}^{c(n)}_l/d\epsilon_s<0$, 
which, from Eq.~(\ref{eq:gammal}), implies that all resonances occurring in
the region of $0<\epsilon_s<B_c(l)$ have positive widths.
Above $B_c(l)$, $\widetilde{\chi}^{c(n)}_l$ is
piecewise monotonically increasing with
$d\widetilde{\chi}^{c(n)}_l/d\epsilon_s>0$, 
implying that all resonances above $B_c(l)$ have 
negative widths.

Resonances of positive width are called shape resonances, consistent
with the standard convention. Resonances of negative width are called
diffraction resonances.
Their distinction can be understood through the concept of the
time delay and the closely related concept of the change of the 
density-of-states due to interaction \cite{wig55,tay06}.
Let $\Delta t_s\equiv \Delta t/s_t$ be the scaled time delay,
where $s_t=\hbar/s_E$ is the time scale associated with the
length scale $\beta_n$.
Let $D_{sl}\equiv D_l/(1/s_E)$ be the scaled change of the 
density-of-states due to interaction.
They are related, and are given by
\begin{equation}
\Delta t_s = 2\pi D_{sl} = 2\frac{d\delta_l}{d\epsilon_s} 
	=\frac{2}{1+K_l^2}\frac{dK_l}{d\epsilon_s}\;.
\end{equation}
It is clear from this equation and Eq.~(\ref{eq:resK}) that a resonance of
positive width corresponds to a time delay \cite{wig55,tay06}
and an enhanced density-of-states,
while a resonances of negative width (diffraction resonance) 
corresponds to a time advance \cite{wig55,Hammer2010} and a reduced 
(negative change) density-of-states. 
For a long-range interaction of the type of $1/r^n$ with $n>3$,
the total number of states is not changed by the interaction, as
is reflected in the Levinson theorem \cite{lev49,joa75}. 
Both the bound states and the shape-resonance states 
can be regarded as states taken from the continuum by the interaction.
The diffraction resonances, which correspond to negative changes of
the density-of-states, give the origin of the bound states and
the shape-resonance states,
namely where in continuum such states come from.

The proceeding discussion on the characterization of scattering 
resonances are applicable to arbitrary $-1/r^n$ potential with
$n>3$. For the specific case of $n=4$,
Figure~\ref{fig:gamma4} gives an illustration of both the width
function and the concept of $B_c(l)$ for partial waves
$l=6$ and 7.
Unlike the width for a shape resonance which can be infinitely
small, the absolute width for a diffraction resonance cannot be
infinitely small. It has a lower limit as required by causality \cite{wig55,Hammer2010}.

\begin{figure}
\includegraphics[width=\columnwidth]{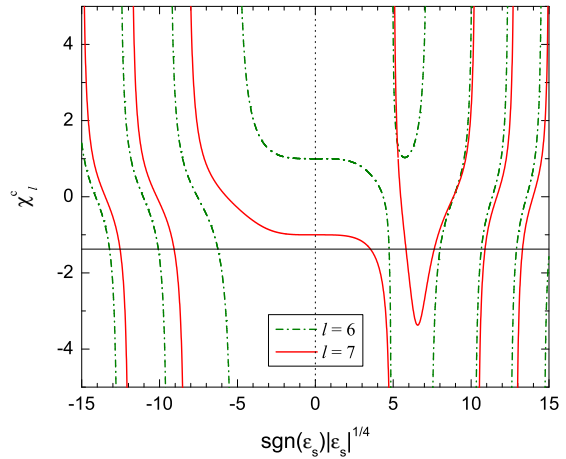}
\caption{(Color online) A closer look at the bound and resonance 
spectra for $l=6$ and $7$ in the $K^c$ representation,
showing the potential existence of two shape resonances for
$l\ge 7$. For $l\le 6$, an energy-independent $K^c$, represented
by a horizontal line, can have only a single crossing point with $\chi^c_l$
within the region of $0<\epsilon_s<B_c(l)$, regardless the value of
$K^c$. For $l=7$ and with proper value of $K^c$, there can
be two such points, corresponding to two shape resonances. 
The horizontal line illustrated at $K^c \approx -1.37638$ corresponds 
to $\mu^c=0.45$ example of Fig.~\ref{fig:pxs4l7}.
\label{fig:usp4chicl67}}
\end{figure}
The QDT formulations for the spectra, together with the concept of
$B_c(l)$, also give the maximum number of shape resonances
that can exist for a particular $l$. They show,
e.g., that the minimum $l$ that can support two shape resonances
through the long-range potential is $l=7$.
Figure~\ref{fig:usp4chicl67} gives a further illustration of 
the relevant concepts, and the difference in behavior between
$l=6$ and $l=7$.
In the $K^c$ and $\mu^c$ representations of the spectra,
the maximum number of shape resonances corresponds to the maximum
number of crossing points between the relevant $\chi_l$ function 
and a straight line that can exist in the region of $0<\epsilon_s<B_c(l)$.
Figure~\ref{fig:usp4chicl67} 
shows that there can only be one such point for $l\le 6$, but
there can be two such points, depending on the short-range parameter,
for $l=7$. Similarly the $\chi_l$ functions for larger $l$ show that the 
minimum $l$ for the existence of 3 shape resonances is $l=19$.
For partial wave $0<l<7$, the increase of the quantum phase $\widetilde{\Phi}^{c(n)}_l$
from the threshold to $B_c(l)$, where it reaches its maximum value,
is less than $\pi$. The phase at $B_0(l)$ (c.f.~Table~\ref{tb:ebins}) is
the same as its value at the threshold.
For $7\le l<19$, the increase of the quantum phase $\widetilde{\Phi}^{c(n)}_l$
from the threshold to $B_c(l)$ is between $\pi$ and $2\pi$, and
the quantum phase at $B_0(l)$ is greater than its value at the threshold
by $\pi$. Similar consideration applies to higher partial waves.

\begin{figure}
\includegraphics[width=\columnwidth]{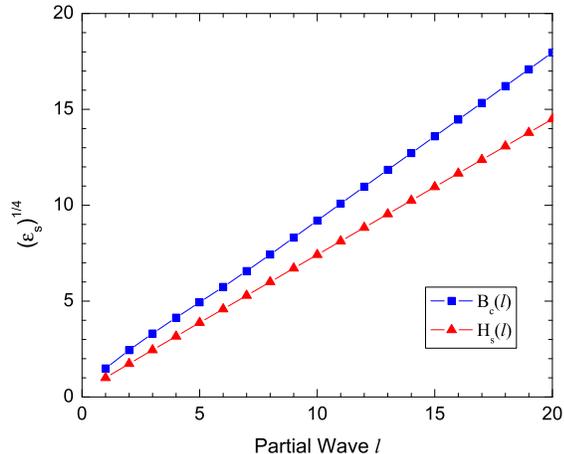}
\caption{(Color online) The critical scaled energies $B_c(l)$ and the scaled 
barrier height $H_s(l)$ for different partial waves. 
It illustrates that both $B_c(l)$ and $H_s(l)$ have $\sim l^4$ type of
$l$ dependence. $B_c(l)$ is always greater than $H_s(l)$, substantially greater
for large $l$, implying that a shape resonance can exist at a substantially
greater energy above the top of the barrier.
\label{fig:Bc4}}
\end{figure}
Figure~\ref{fig:Bc4} shows the $B_c(l)$ for a large number of
partial waves, together with the scaled barrier height for
a $-1/r^4$ potential given by 
$
H_{s}(l) = [l(l+1)]^2/4
$.
It illustrates that (a) both $B_c(l)$ and $H_s(l)$ have $\sim l^4$ type of
$l$ dependence, and (b) $B_c(l)$ is always greater than $H_s(l)$, substantially greater
for large $l$, implying that a shape resonance can exist at a substantially
greater energy above the top of the barrier.

With the introduction and the interpretation of the universal
width function, we can now provide a more complete characterization of
the five resonances in Fig.~\ref{fig:pxs4l7}.
The resonance positions can be predicted from any of the three formulations
of the spectrum in Sec.~\ref{sec:qdt}, and the widths are evaluated
from the universal width function.
In partial wave $l=7$ with $\mu^c(l=7)=0.45$, there are two shape resonances located at
$\epsilon_s = 157.041$, and $\epsilon_s = 1148.23$,
with scaled widths of $1.099\times 10^{-3}$ and $1.851 \times 10^3$,
respectively. They are shape resonances with positive widths.
The scaled barrier height for $l=7$ is $H_s(l=7)=784$. The narrow shape resonance
is substantially below the barrier, and the broad shape resonance
is above the barrier and below the $B_c(l=7)\approx 1863.58$.
The other three resonances are located at
$\epsilon_s = 3390.55, 14082.9$, and 31297,
with scaled widths of $-6.220\times 10^3$, $-8.542\times 10^3$,
and $-1.362\times 10^4$, respectively.
They are diffraction resonances with negative widths.
We emphasize that without the concept of the diffraction resonance, 
not all the features in
Figure~\ref{fig:pxs4l7} would be accounted for.
We further emphasize that from a pure mathematical point of view, 
any attempt to represent the $K$ matrix,
namely the $K_l$ function, as a smooth background function 
plus a set of poles over the entire positive energy range 
can never ignore poles associated with diffraction resonances. 
The only difference between these poles and those associated with
shape resonances is that they have a residue of a different sign.

\subsection{Universal spectral properties}
\label{sec:usp4}

\begin{figure}
\includegraphics[width=\columnwidth]{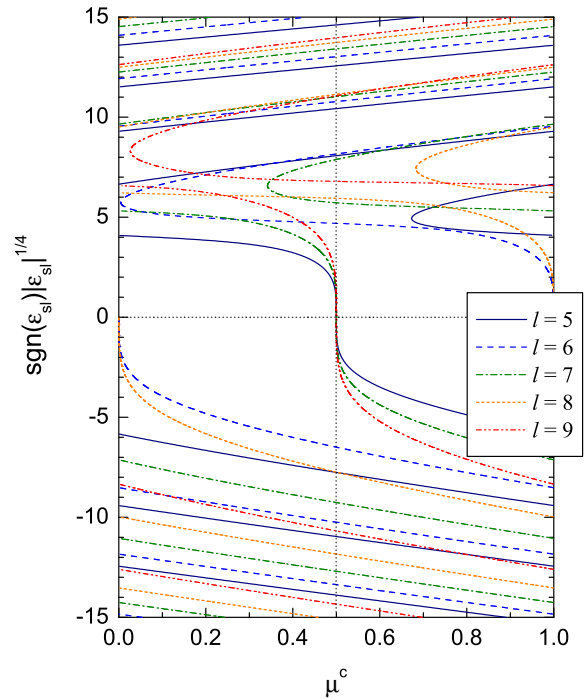}
\caption{(Color online) The universal spectra for quantum systems with 
a $-1/r^4$ type of long-range interaction, versus $\mu^c(l)$ for $l=5$-$9$.
The applicability of each individual curve, which gives the spectrum
of a particular partial wave $l$ from the corresponding short-range
parameter $\mu^c(l)\equiv \mu^c(\epsilon=0,l)$, requires only the 
energy independence of the parameter.
For systems such as ion-atom,
for which $\mu^c(l)$ is approximately independent of $l$,
a single $\mu^c(l=0)$ gives the entire spectra for all $l$. A similar figure
for lower partial waves can be found in Ref.~\cite{gao10a}.
\label{fig:usp4emuc}}
\end{figure}
\begin{figure}
\includegraphics[width=\columnwidth]{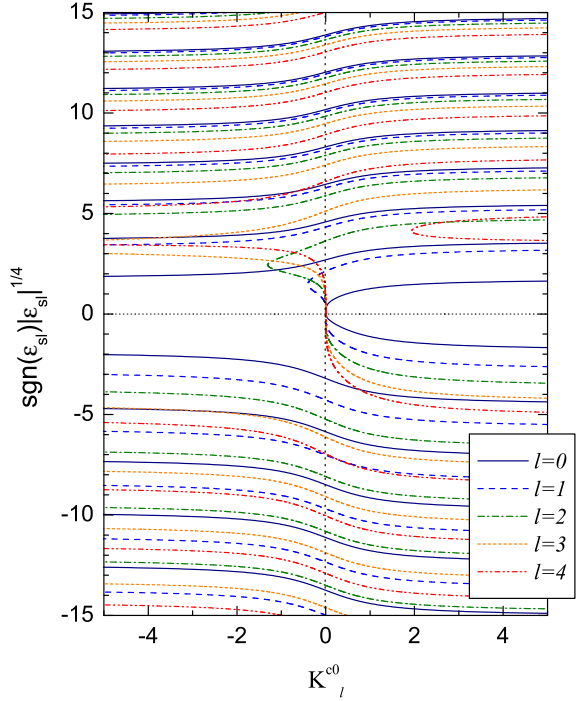}
\caption{(Color online) The universal spectra for quantum systems with 
a $-1/r^4$ type of long-range interaction, versus 
$K^{c0}_l\equiv K^{c0}_l(\epsilon=0)$ for $l=0$-$4$.
Since $K^{c0}_l$ is related to the generalized scattering
length by Eq.~(\ref{eq:Kc0ga}), this set of curves also represent
the spectra as a function of the inverse of a 
reduced generalized 
scattering length, $1/(\widetilde{a}_{l}/\bar{a}^{(4)}_l)$. 
\label{fig:usp4eKc0}}
\end{figure}
Under the assumption that the
short-range parameter, $K^c(\epsilon,l)$, $\mu^c(\epsilon,l)$, or $K^{c0}_l(\epsilon)$,
being independent of energy, they are constants that can be taken to be their values at
the zero energy.
Defining $K^c(l)\equiv K^c(\epsilon=0,l)$, $\mu^c(l)\equiv\mu^c(\epsilon=0,l)$, 
and $K^{c0}_l\equiv K^{c0}_l(\epsilon=0)$,
the equations for the spectra can be solved (inverted) to give both the scaled binding energies
and the scaled resonance positions, $\epsilon_{sl}$, as a function of a short-range
parameter, $K^c(l)$, $\mu^c(l)$, or $K^{c0}_l$.
Figures~\ref{fig:usp4emuc} and \ref{fig:usp4eKc0} illustrate the results
versus $\mu^c(l)$ and $K^{c0}_l$, respectively.
They are the $-1/r^4$ equivalents of the Rydberg formula for the
Coulomb interaction, generalized to include also the resonance
spectrum.
They greatly generalize the well-known result
of the effective range theory \cite{sch47,bla49,bet49}, 
$\epsilon_{sl=0}\sim -(a_{l=0}/\beta_4)^{-2}$, for the $s$ wave
least-bound state, to more deeply bound states,
to all partial waves, and to resonance positions.
The representation in terms of $\mu^c(l)$, Fig.~\ref{fig:usp4emuc}, is the most
convenient for a systematic understanding of ion-atom spectra,
for which the $\mu^c(l)$ has the additional characteristic of
being approximately partial-wave-independent.
The representation in terms of $K^{c0}_l$, Fig.~\ref{fig:usp4eKc0}, is most convenient
for illustrating the dependence of the spectra on the scattering
length or the generalized scattering length \cite{gao09a}, as we now
explain.

For any potential of the type of $-1/r^n$ with $n>3$, the $s$ wave 
scattering length is well defined (at zero energy), and is related to
the other short-range parameters at zero energy through
$K^c(\epsilon=0,l=0)$ by \cite{gao03,gao04b}
\begin{equation}
a_{l=0}/\beta_n = \left[b^{2b}\frac{\Gamma(1-b)}{\Gamma(1+b)}\right]
    \frac{K^c(0,0) + \tan(\pi b/2)}{K^c(0,0) - \tan(\pi b/2)} \;,
\label{eq:a0sKc}
\end{equation}
where $b=1/(n-2)$. For $n=4$, this relation reduces to \cite{LG12}
\begin{equation}
a_{l=0}/\beta_4 = \frac{K^c(0,0) + 1}{K^c(0,0) - 1} \;.
\label{eq:a0sKc4}
\end{equation}
Combining it with Eq.~(\ref{eq:Kc0Kc}),
we have
\begin{equation}
K^{c0}_{l=0}(\epsilon=0) = \frac{1}{a_{l=0}/\beta_4} \;.
\label{eq:Kc0a0}
\end{equation}
Equation~(\ref{eq:Kc0a0}) means that at least for the $s$ wave, 
Fig.~\ref{fig:usp4eKc0} 
gives in fact a representation of the spectrum as a function
of the inverse of a reduced scattering length.

Combining Eqs.~(\ref{eq:Kc0muc}) and (\ref{eq:Kc0a0}) gives
\begin{equation}
a_{l=0}/\beta_4 = \cot[\pi\mu^c(\epsilon=0,l=0)] \;,
\label{eq:a0smuc}
\end{equation}
which relates the $s$ wave scattering length to the $s$ wave 
quantum defect evaluated at the zero energy, with the range of
$0\le \mu^c(\epsilon=0,l=0)<0.5$ corresponding to positive
$s$ wave scattering length, 
$0.5<\mu^c(\epsilon=0,l=0)<1$ corresponding to negative
$s$ wave scattering length, and $\mu^c(\epsilon=0,l=0)=0.5$ corresponding
to the zero $s$ wave scattering length.

For other partial waves, the scattering length is not defined in the
conventional sense \cite{oma61,lev63}. However, 
in an upcoming publication, we will show that the concept of
scattering length can be generalized to all partial waves
through the QDT expansion \cite{gao09a} for the $-1/r^4$ potential. 
This generalized scattering
length is related to $K^{c0}_l(\epsilon)$ at zero energy by
\begin{equation}
K^{c0}_l(\epsilon=0) = \frac{1}{\widetilde{a}_l/\bar{a}^{(4)}_{l}} \;,
\label{eq:Kc0ga}
\end{equation}
where $\bar{a}^{(4)}_{l}=\bar{a}^{(4)}_{sl}\beta_4^{2l+1}$ is called 
the mean scattering length for the $-1/r^4$ potential in partial wave $l$, with
\begin{eqnarray}
\bar{a}^{(4)}_{sl} &=& \frac{\pi^2}{2^{4l}(2l+1)^2[\Gamma(l+1/2)]^2} \nonumber\\
	&=& \frac{(2l+1)^2}{[(2l+1)!!]^4} \;,
\end{eqnarray}
being what we call the scaled mean scattering length for the $-1/r^4$ 
potential in partial wave $l$ \cite{gao11a}. 
Thus $K^{c0}_l$ at the zero energy is the inverse of a reduced
generalized scattering length, not only for the $s$ wave, where Eq.~(\ref{eq:Kc0ga})
reduces to Eq.~(\ref{eq:Kc0a0}), but for all partial waves.
Figure~\ref{fig:usp4eKc0} gives the spectrum as a function
of this inverse reduced generalized scattering length for all partial waves.
The generalized scattering length is related to the quantum defect
evaluated at the zero energy, by
\begin{equation}
\widetilde{a}_l/\bar{a}^{(4)}_{l} = \left\{
	\begin{array}{ll}
	\cot[\pi\mu^c(\epsilon=0,l)] \;, & l=\text{even} \\
	-\tan[\pi\mu^c(\epsilon=0,l)] \;,& l=\text{odd}
	\end{array} \right.\;.
\label{eq:gamuc}	
\end{equation}
It reduces to Eq.~(\ref{eq:a0smuc}) for the $s$ wave.
\begin{figure}
\includegraphics[width=\columnwidth]{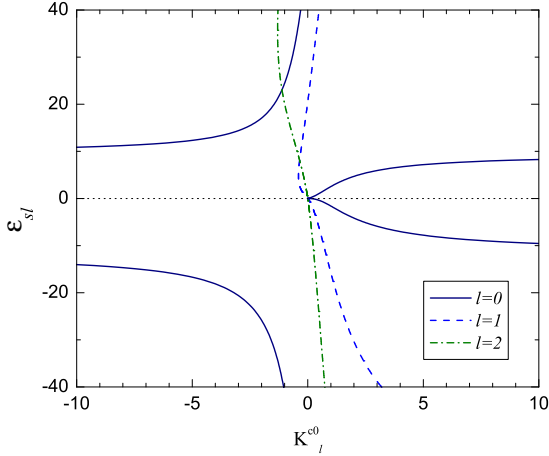}
\caption{(Color online) A magnified view of a small region 
of Figure~\ref{fig:usp4eKc0} around the threshold, showing the
bound state energy and the resonance position $\epsilon_{sl}$ on
a linear scale. It is easier to observe on this figure
the expected $\epsilon_{sl=0}\sim -(a_{l=0}/\beta_4)^{-2}$
behavior for the $s$ wave bound state energy around the threshold.
\label{fig:usp4esKc0}}
\end{figure}
Figure~\ref{fig:usp4esKc0} is a magnified view of a small region 
of Figure~\ref{fig:usp4eKc0} around the threshold.
It is used to illustrate the expected $\epsilon_{sl=0}\sim -(a_{l=0}/\beta_4)^{-2}$
behavior for the $s$ wave bound state energy around the threshold.

All universal properties discussed up to this point assume only
the energy independence of the short-range parameter.
For single-channel ion-atom interactions, the spectra follow
closely a set of universal behaviors that are derived under the
further assumption that the short-range parameter, 
$K^c(\epsilon,l)$ or $\mu^c(\epsilon,l)$ [but not $K^{c0}_l(\epsilon)$],
is not only independent of energy, but also independent of the
partial wave $l$ \cite{gao01,gao04b,LG12}. In this case, interaction in different partial
waves become related. Other than a single overall energy scaling
factor $s_E$, every aspect of
ion-atom interaction, including the entire
rovibrational spectrum and all scattering 
properties in all partial waves, can be determined from a single 
parameter \cite{gao01,LG12}.

Some of the consequences on the spectra that emerge for an $l$ independent
$\mu^c$ or $K^c$ have been discussed in Ref.~\cite{gao04b} 
in the general context of an arbitrary $-1/r^n$ potential with $n>2$.
Figure~\ref{fig:usp4emuc}, together with Figure~1 of Ref.~\cite{gao10a},
illustrates explicitly how such properties manifest themselves
in the spectra for the $-1/r^4$ potential. Specifically, they show explicitly the
following characteristics of a single-channel ion-atom system.
(a) Having a quasibound $s$ state right at the threshold
means having a bound state right at the threshold for all even partial
waves with $l=0+(n-2)j=2j$ where $j$ is a nonnegative integer. Similarly, having a $p$ wave bound state
right at the threshold means having a bound state right at the threshold 
for all odd partial waves with $l=1+(n-2)j=2j+1$.
(b) The least-bound state for a single-channel ion-atom system 
is either an $s$ state or a $p$ state,
depending on the quantum defect \cite{gao04b}.
For systems with $0\le \mu^c<0.5$, corresponding to 
positive $s$ wave scattering lengths, the least bound state is
an $s$ state. For systems with $0.5< \mu^c<1$, corresponding to 
negative $s$ wave scattering lengths, the least bound state is
an $p$ state.
(c) Systems with a quantum defect smaller than but close to 0.5,
correspond to a small positive $s$ wave scattering length,
have a set of narrow shape resonances in odd partial waves. 
Systems with a quantum defect smaller than but close to 1.0,
corresponding to a large negative $s$ wave scattering length,
have a set of narrow shape resonances in even partial waves.

\subsection{Universal total elastic cross section for single-channel ion-atom scattering}

\begin{figure}
\includegraphics[width=\columnwidth]{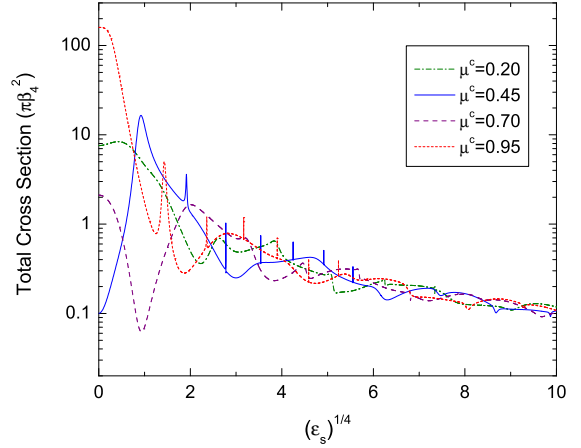}
\caption{(Color online) Illustrations of universal single-channel ion-atom total elastic
cross section. Different
systems differ only in scaling as determined by the atomic polarizability,
and a single quantum defect $\mu^c$, which can be determined from a single
experimental measurement of either a resonance position or a
bound state energy of a vibrationally highly-excited state. 
Note the existence of many narrow shape resonances, similar to that
illustrated for $l=7$ in Fig.~\ref{fig:pxs4l7}, for $\mu^c$ smaller than but close
to either 0.5 or 1.
\label{fig:utxs4muces}}
\end{figure}
For single-channel ion-atom interactions,
having a single parameter being able to describe multiple partial
waves implies that not only a partial cross section would follow a
universal behavior, but also the total cross section \cite{LG12}.
Figure~\ref{fig:utxs4muces} illustrates the universal behavior 
of the total cross section that is followed in varying degrees 
by all single-channels ion-atom systems.
They include all $^1$S+$^1$S type of ion-atom systems 
(corresponding to a single $^1\Sigma$ molecular state), such as a 
Group IA ion (e.g. Li$^+$) or Group IIIA ion (e.g. Al$^+$)
with Group IIA atoms (e.g. Mg) or Group VIIIA atoms (e.g. Ar).
It is also applicable to all $^1$S+$^2$S type of ion-atom interactions
(a single $^2\Sigma$ molecular state) provided
that the atoms involved are dissimilar, namely has different atomic number $Z$,
and the ionization potentials of the atoms involved are such that there
is either no charge transfer channel open at low energies or that the transfer
cross section is small. The examples
include interactions of Group IIA ions with Group VIIIA atoms, such as Mg$^+$+Ar,
Group IIA ions with Group IIA atoms of a difference species, such as Mg$^+$+Ca,
Group IIIA ion with Group IA atoms,
Group IA atoms with Group IA ion of a different species, such as Li$^+$+Na.
For sufficiently small energies around the threshold,
\textit{all such systems} differ only in scaling as determined by the 
atomic polarizability, and a single quantum defect.

Figure~\ref{fig:utxs4muces} further illustrates the potential existence of extremely narrow
shape resonances, many of which would almost definitely be missed
on any calculation performed on a finite energy mesh.
In order not to miss them, one needs first to identify their existence and locations using
the resonance spectrum and to calculate their widths using the width
function. A special mesh of energies are then generated around the
narrow resonance locations using the width information. 
Cross section calculations performed on the special mesh are
added to those on the regular mesh to give the results of
Figure~\ref{fig:utxs4muces}.

The single-channel universal behavior for ion-atom interaction has
been verified through comparison with numerical calculations \cite{LG12}.
The energy range of applicability of the universal behavior, as measure in units of $s_E$, 
is determined by the length scale separation,
more specifically by $\beta_6/\beta_4$, where
$\beta_6=(2\mu C_6/\hbar^2)^{1/4}$ is the length scale
associated with the $-C_6/r^6$ term of the ion-atom potential,
as in $V(r)\sim -C_4/r^4-C_6/r^6$.
As illustrated in Ref.~\cite{LG12},
this range of energy is typically $10^5s_E$ if the atom 
involved is an alkali-metal atom.
Beyond this range, accurate QDT calculations will require
incorporation of the $l$ dependence of the
short-range parameter, as will be illustrated elsewhere.

\section{Discussions}
\label{sec:discuss}

All equations of this work are formally applicable to both ion-atom and
electron-atom or positron-atom interactions. There are however significant
differences in how they are actually used in those different
contexts, as already mentioned throughout this work. 
We briefly summarize some of the key differences here for the
sake of clarity and future applications.

For ion-atom interactions, $K^c$ depends very weakly on energy
on a scale of $s_E$, due to the fact the $\beta_4$ is much greater
than other length scales in the system \cite{gao10a,LG12}. 
It also depends weakly on the partial wave $l$,
a feature that is very important for its application beyond
the ultracold regime \cite{LG12}. 
For an accurate characterization of low-energy ion-atom interaction,
the key difficulty is the sensitive dependence of the short-range
parameter and the scattering length on the short-range 
potential \cite{gao96,LG12}.
This difficulty is overcome in the QDT formulation through
direct determination of such parameters from one or a few
measurements of either the binding energy of a loosely-bound molecular
ion \cite{gao98b,gao01}, or the resonance position.
This is simply done by using the representations of the spectrum
in reverse. For instance, knowing a single bound state energy
$\epsilon_{l}$ in one particular partial wave $l$, we have from Eq.~(\ref{eq:ubsp})
\begin{equation}
K^c(\epsilon_{l},l) = \chi^{c(n)}_l(\epsilon_{sl}) \;.
\end{equation} 
In other words, the $\chi^{c(n)}_l$ function evaluated at 
$\epsilon_{sl}=\epsilon_l/s_E$
gives the value of $K^c$ in partial wave $l$ at $\epsilon_l$.
Since $K^c$ depends weakly on both the energy and the partial wave,
this single value is already sufficient to predict, in a region around the
threshold, all other bound states and scattering properties in
all partial waves.
The generalized scattering length for any partial wave,
which is but one special scattering property at zero energy,
can be obtained in a number of ways.
For example, from $K^c(\epsilon=0,l)\approx K^c(\epsilon_l,l)$, we obtain
$\mu^c(\epsilon=0,l)$, from which the generalized 
scattering length for partial wave $l$ can be obtained from
Eq.~(\ref{eq:gamuc}). This prediction only assumes the energy independence of
$K^c$ and $\mu^c$. Making further use of their $l$ independence,
Eq.~(\ref{eq:gamuc}) would give the generalized scattering 
lengths for all partial waves.
This procedure works the same for a resonance position, with the
only difference being the $\chi^{c(n)}_l$ in the above equation
being replaced by $\widetilde{\chi}^{c(n)}_l$.
One or a few more experimental data points for bound state energy 
and/or resonance position
would enable the extraction the $C_4$
coefficient (thus the atomic polarizability),
in a procedure parallel to that of Ref.~\cite{gao01}
for the $C_6$ coefficient.
They would also enable a more accurate representation of 
$K^c(\epsilon,l)$, and thus a more accurate representation of the
spectra and scattering properties, over a wider range of energies
and partial waves. 

For electron-atom interaction, we lose the weak dependence of 
$K^c$ on $l$ \cite{gao01,gao08a}.
One needs one short-range parameter for each partial wave.
$K^c$ can also have a more significant energy dependence over a 
scale of $s_E$, at least for atoms in their ground states,
for which the $\beta_4$ is not much greater than the size of
an atom due to the small mass of the electron.
These ``complications'' are countered by the fact that much 
fewer partial waves would contribute at a fixed energy.
For example, electron-alkali interaction is dominated by
$s$ wave scattering even at the room temperature of $\epsilon/k_B\sim 300$ K,
where alkali interactions with alkali ions 
would have required hundreds of partial waves.
For an accurate description of low-energy
electron-atom interaction, the key difficulty and focus is instead
on the accurate characterization of the energy dependence of the short
range parameters for the first few partial waves \cite{fab86}.

In the context of electron interaction with a ground-state atom, 
the energy bins, $B_{-1}(l)$,
translate into upper bounds for electron affinities.
For example, $B_{-1}$ for the $s$ wave translates
into a upper bound for electron affinities $EA$ 
for all alkali-metal atoms and hydrogen, as
$EA \le 105.8078 s_E$,
which has been verified using the data in 
Ref.~\cite{hot85}.

\section{Conclusions}
\label{sec:conclusion}

In conclusion, we have presented a detailed QDT formulation 
for the $-1/r^4$ potential.
The concepts of resonance spectrum, diffraction resonance,
and universal width function, 
are discussed in detail in the general context of $-1/r^n$ potential
and illustrated for the $-1/r^4$ potential.
The theory provides a solid foundation for a systematic 
understanding of charge-neutral quantum systems that 
include ion-atom, ion-molecule, 
electron-atom, and positron-atom interactions.
For example, the QDT description of narrow shape resonances 
gives hope for a better understanding of
their effects on chemical reactions \cite{cha10}, 
on radiative association \cite{bar06}, and on thermodynamics.

This presentation of QDT is the first general presentation since
the works of Refs.~\cite{gao08a,gao10a}, and includes ingredients
not found in earlier QDT formulations for 
$1/r^6$ \cite{gao98b,gao01} and $1/r^3$ \cite{gao99a,gao99b} potentials.
It should be clear from this work that the QDTs for $-1/r^6$
and $-1/r^3$ potentials can be recast, extended, and understood
in a similar manner as the $-1/r^4$ theory presented here.

In Ref.~\cite{LG12}, we have demonstrated how this version of QDT,
even in its simplest parametrization, provide a accurate
characterization of ion-atom interaction over a much greater
range of energies (by 5 orders of magnitude) than the effective-range 
expansion \cite{oma61}, using the same parameters.
In subsequent publications, we will show how an improved
parametrization can provide a \textit{quantitative} description
over an even greater range of energies
and for both single-channel and multichannel processes.
The existence of such a systematic theory, together with
similar theories for neutral-neutral quantum systems \cite{gao05a,gao08a},
offer a prospect, in our view, for new classes of
quantum theories for few-body and many-body systems, including
systems of mixed species, that will be applicable over
a much greater range of temperatures and densities than
theories based on the effective range descriptions
of interactions \cite{sch47,bla49,bet49,oma61}.

\begin{acknowledgments}
I thank Li You, Meng Khoon Tey, Ming Li, and Haixiang Fu
for helpful discussions and for careful reading of the manuscript.
This work was supported in part by NSF.
\end{acknowledgments}

\appendix
\section{The determination of the characteristic exponent}
\label{sec:charexp4}
The characteristic exponent, $\nu$, as its name implies,
plays a central role in the theory of Mathieu 
class of functions \cite{olv10}, and similarly in 
solutions for $1/r^6$ and $1/r^3$ potentials \cite{gao98a,gao99a}. 
In all these cases, it characterizes the nature of the nonanalytic behavior
of the solutions at $\epsilon=0$ in the energy space, and at 
both essential singularities, $r=0$ and $r=\infty$, in the coordinate space.
In the context of solutions for Schr\"odinger equations, 
$\nu$ is a function of a scaled energy for each partial wave. 
Once $\nu$ is determined, every other
aspect of the solution follows in a straightforward manner.

It is known that the characteristic exponent for Mathieu 
class of functions can be determined using two different 
methods \cite{mor53v1}. One is as the root of
a Hill determinant.
The other is as the root of a characteristic function.
In the Hill determinant formulation, it is a solution of
\begin{equation}
{\cal D}^H_l(\nu;\epsilon_s) = 0 \;,
\label{eq:DHeq0}
\end{equation}
where ${\cal D}^H_l(\nu;\epsilon_s)$ is the Hill determinant
corresponding to the three-term recurrence relation,
Eq.~(\ref{eq:mathieurec}),
\begin{equation}
{\cal D}^H_l\equiv\det
\left(
\begin{array}{ccccccc}
\ddots & \vdots & \vdots & \vdots & \vdots & \vdots &  \\
\dots  & 1      & h_{-2} & 0      & 0      & 0      & \dots \\
\dots  & h_{-1} & 1      & h_{-1} & 0      & 0      & \dots \\
\dots  & 0      & h_{0}  & 1      & h_{0}  & 0      & \dots \\
\dots  & 0      & 0      & h_{1}  & 1      & h_{1}  & \dots \\
\dots  & 0      & 0      & 0      & h_{2}  & 1      & \dots \\
       & \vdots & \vdots & \vdots & \vdots & \vdots & \ddots \\
\end{array}
\right) \;,
\end{equation}
where $h_m$ is defined by Eq.~(\ref{eq:hm}).

In the characteristic function method, $\nu$ is a
solution of
\begin{equation}
\Lambda_l(\nu;\epsilon_s) = 0\;,
\end{equation}
where
\begin{equation}
\Lambda_l(\nu;\epsilon_s) \equiv (\nu^2-\nu_0^2)
	- \epsilon_s[\bar{Q}(\nu)+\bar{Q}(-\nu)]\;,
\label{eq:charact}
\end{equation}
is the characteristic function, with 
$\bar{Q}(\nu)$ defined in terms of the $Q(\nu)$ function
[cf. Eq.~(\ref{eq:Qcf4})] by
\begin{equation}
\bar{Q}(\nu)= \frac{1}{(\nu+2)^2-\nu_0^2}Q(\nu) \;.
\label{eq:Qbar}
\end{equation}

It can be shown that the Hill determinant and the characteristic
function are related by
\begin{equation}
{\cal D}^H_l(\nu,\epsilon_s) = \frac{1}{(\nu^2-\nu_0^2)
	C_{\epsilon_s l}(\nu)C_{\epsilon_s l}(-\nu)}
	\Lambda(\nu;\epsilon_s) \;.
\label{eq:DHlam}	
\end{equation}
This relationship, which we have not found elsewhere,
not only makes it immediately clear that the two approaches
to $\nu$ are equivalent, but also provides an efficient
method for the evaluation of the Hill determinant and
therefore the characteristic exponent.

Due to the special characteristics of a Hill determinant \cite{whi96},
the solution of Eq.~(\ref{eq:DHeq0}) can be found
through the evaluation of ${\cal D}^H_l(\nu;\epsilon_s)$
at a single $\nu$ such as $\nu=0$.
Defining
${\cal H}_l(\epsilon_s)\equiv {\cal D}^H_l(\nu=0;\epsilon_s)$,
we have from Eq.~(\ref{eq:DHlam})
\begin{equation}
{\cal H}_l(\epsilon_s) = \frac{1}{\nu_0^2[C_{\epsilon_s l}(0)]^2}
	\left[\nu_0^2+\frac{2\epsilon_s}{4-\nu_0^2}Q(0)\right] \;.
\end{equation}

From ${\cal H}_l$, the $\nu$, as a function of the scaled energy, can
be found as the solutions of \cite{hol73,gao10a}
\begin{equation}
\sin^2(\pi\nu/2)= {\cal H}_l(\epsilon_s)/2 \;.
\end{equation}
For example, for ${\cal H}_l<0$ or ${\cal H}_l>2$, $\nu = \nu_r+i\nu_i$ is complex, 
with its imaginary part $\nu_i$ given by
\begin{eqnarray}
\nu_i &=& \frac{1}{\pi}\cosh^{-1}\left(|1-{\cal H}_l|\right)\;,\\
	&=& \frac{1}{\pi}\ln\left[|1-{\cal H}_l|
	+\sqrt{(1-{\cal H}_l)^2-1}\right] \;.
\end{eqnarray}
Its real part is given by
\begin{equation}
\nu_r = \left\{
\begin{array}{ll}
l\;,  & l=\mathrm{even} \\
l+1\;,& l=\mathrm{odd}
\end{array}
\right. \;,
\end{equation}
for ${\cal H}_l<0$, and by
\begin{equation}
\nu_r = \left\{
\begin{array}{ll}
l+1\;,  & l=\mathrm{even} \\
l\;,    & l=\mathrm{odd}
\end{array}
\right. \;,
\end{equation}
for ${\cal H}_l>2$. The real part of $\nu$
is defined within a range of 2. All $\nu+2j$, where $j$ is an 
integer, are equivalent.

\section{Comparison of notations}

In presenting mathematical results for modified Mathieu functions,
we have adopted notations derived from our earlier solutions
of $1/r^6$ \cite{gao98a} and $1/r^3$ \cite{gao99a} potentials,
to emphasize their structural similarities.

Prior to recent works \cite{gao10a,idz11},
the most detailed study of modified Mathieu functions, in a domain
most relevant to the solutions of the Schr\"{o}dinger equation for $-1/r^4$
potential, has been the work of Holzwarth  \cite{hol73}.
To make it easier to relate this work to earlier works \cite{hol73,khr93},
we summarize, in Table~\ref{tb:notation}, the correspondence between our
notations and notations of Holzwarth \cite{hol73}.

\begin{table}
\caption{Comparison of notations of Holzwarth \cite{hol73}
and our notations. \label{tb:notation}}
\begin{ruledtabular}
\begin{tabular}{rrrr}
Holzwarth\footnotemark[1] & Our notation & Holzwarth\footnotemark[1] & Our notation     \\
\hline
$\tau$     & $\nu$                  & $K^-$        & $G_{\epsilon_s l}(-\nu)$     \\
$\Delta^l(0)$     & ${\mathcal H}_l$    & $K^-/K^+$        & $M_{\epsilon_s l}$     \\
$K^+$     & $G_{\epsilon_s l}(+\nu)$ & &
\end{tabular}
\end{ruledtabular}
\footnotetext[1]{Ref.~\cite{hol73}}
\end{table}

\bibliography{bgao,twobody,eatom,ionAtom,ionChem,Rydberg,molMol,Fesh,coldMolion,numerical}

\end{document}